\documentclass[aps,prl,reprint,superscriptaddress,amssymb,longbibliography,color,tabulary,floatfix]{revtex4-1}

\usepackage{blindtext}
\usepackage{graphicx}
\usepackage{natbib}
\usepackage{siunitx}
\usepackage{textcomp}
\usepackage{gensymb}
\usepackage{multirow}
\usepackage[dvipsnames]{xcolor}
\usepackage{hyperref}
\hypersetup{colorlinks=true,linkcolor=blue,filecolor=magenta,urlcolor=cyan}
\usepackage[nameinlink, noabbrev]{cleveref}

\newcommand{\orm}{\textit{Observatorio del Roque de los Muchachos\ }}
\newcommand{\NOT}{\textit{Nordic Optical Telescope}}
\newcommand{\tng}{\textit{Telescopio Nazionale Galileo}\ }
            \newcommand{\wht}{\textit{William Herschel Telescope}\ }
            \newcommand{\source}{$\mathcal{S}$}
\newcommand{\alice}{$\mathcal{A}$}
\newcommand{\bob}{$\mathcal{B}$}
\newcommand{\ket}[1]{\left|#1\right\rangle} %erzeugt einen ket: |M>

\newcommand{\yr}{yr}
\newcommand{\Vexp}{\mathcal{V}}

\newcommand{\aap}{Astron. Astrophys.}
\newcommand{\aj}{Astron. J.}
\newcommand{\apjl}{Astrophys. J. Lett.}
\newcommand{\jcap}{J. Cosmol. Astropart. Phys}

\newcommand{\pasj}{Publ. Astron. Soc. Japan}

\begin{document}
	\title{Cosmic Bell Test using Random Measurement Settings from High-Redshift Quasars}

\newcommand{\iqoqi}{Institute for Quantum Optics and Quantum Information (IQOQI), Austrian Acadamy of Sciences, Boltzmanngasse 3, 1090 Vienna, Austria}
\newcommand{\MIT}{Department of Physics, Massachusetts Institute of Technology, Cambridge, Massachusetts 02139, USA}
\newcommand{\vcq}{Vienna Center for Quantum Science \& Technology (VCQ), Faculty of Physics, University of Vienna, Boltzmanngasse 5, 1090 Vienna, Austria}
\newcommand{\nudt}{School of Computer, NUDT, 410073 Changsha, China}
\newcommand{\harvey}{Department of Physics, Harvey Mudd College, Claremont, California 91711, USA}
\newcommand{\uc}{Center for Astrophysics and Space Sciences, University of California, San Diego, La Jolla, California 92093, USA}
\newcommand{\ing}{Isaac Newton Group, Apartado 321, 38700 Santa Cruz de La Palma, Spain}
\newcommand{\inaf}{Fundaci\'{o}n Galileo Galilei---INAF, 38712 Bre\~na Baja, Spain}

\author{Dominik Rauch}
\email{dominik.rauch@oeaw.ac.at}
\author{Johannes Handsteiner}
\author{Armin Hochrainer}
\affiliation{\iqoqi}
\affiliation{\vcq}
\author{Jason Gallicchio}
\affiliation{\harvey}
\author{Andrew S. Friedman}
\affiliation{\uc}
\author{Calvin Leung}
\affiliation{\iqoqi}
\affiliation{\vcq}
\affiliation{\harvey}
\affiliation{\MIT}
\author{Bo Liu}
\affiliation{\nudt}
\author{Lukas Bulla}
\author{Sebastian Ecker}
\author{Fabian Steinlechner}
\author{Rupert Ursin}
\affiliation{\iqoqi}
\affiliation{\vcq}
\author{Beili Hu}
\affiliation{\harvey}
\author{David Leon}
\affiliation{\uc}
\author{Chris Benn}
\affiliation{\ing}
\author{Adriano Ghedina}
\author{Massimo Cecconi}
\affiliation{\inaf}
\author{Alan H. Guth}
\author{David I. Kaiser}
\email{dikaiser@mit.edu}
\affiliation{\MIT}
\author{Thomas Scheidl}
\author{Anton Zeilinger}
\email{anton.zeilinger@univie.ac.at}
\affiliation{\iqoqi}
\affiliation{\vcq}
	
	\begin{abstract}
 In this paper we present a cosmic Bell experiment with polarization-entangled photons, in which measurement settings were determined based on real-time measurements of the wavelength of photons from high-redshift quasars, whose light was emitted billions of years ago; the experiment simultaneously ensures locality. Assuming fair sampling for all detected photons, and that the wavelength of the quasar photons had not been selectively altered or previewed between emission and detection, we observe statistically significant violation of Bell's inequality by $9.3$ standard deviations, corresponding to an estimated $p$ value of $\lesssim 7.4 \times 10^{-21}$. This experiment pushes back to at least $\sim$7.8\,Gyr ago the most recent time by which any local-realist influences could have exploited the ``freedom-of-choice" loophole to engineer the observed Bell violation, excluding any such mechanism from $96\%$ of the space-time volume of the past light cone of our experiment, extending from the big bang to today.
	\end{abstract}
	
	\maketitle

\paragraph{Background.}
To Erwin Schr\"odinger, entanglement was ``{\it the} characteristic trait of quantum mechanics, the one that enforces its entire departure from classical lines of thought" \cite{Schrodinger1935a}.
He referred to an analysis by Einstein, Podolsky, and Rosen (EPR) \cite{Einstein1935}, regarding the quantum-mechanical predictions for perfect correlations in certain quantum systems.
EPR made two assumptions explicit. Regarding locality, they wrote: ``Since at the time of measurement the two systems no longer interact, no real change can take place in the second system in consequence of anything that may be done to the first system." They also articulated a ``reality criterion": ``If, without in any way disturbing a system, we can predict with certainty (i.e., with probability equal to unity) the value of a physical quantity, there exists an element of physical reality corresponding to this physical quantity." In the light of these two assumptions and their analysis of a particular two-particle state, EPR concluded that quantum mechanics is incomplete. While the EPR reasoning is logically unassailable, Niels Bohr pointed out that the EPR assumptions need not hold for quantum observations \cite{Bohr1935}.

This discussion had laid dormant for several decades, but in 1964 John Stewart Bell demonstrated that a complete theory based on the EPR premises makes predictions that are in conflict with those of quantum mechanics \cite{Bell1964,BellSpeakable}. In such local-realist theories, it is assumed that every individual system carries its own set of properties prior to measurement, which are presumed to be independent of any possible influence from outside its past light cone. Bell concluded that in a local-realist theory the strength of correlations among measurements on different particles' properties is limited and smaller than the predictions of quantum physics. This is expressed by
Bell's inequality.

With Bell's result, a question that previously had been dismissed as ``merely philosophical"
became experimentally testable. In 1969, Clauser, Horne, Shimony, and Holt (CHSH) published their inequality as an experimentally accessible variant of Bell's version \cite{Clauser1969}. The idea was to measure the four probabilities $p(A,B \vert a_i, b_j)$ of measurement results $A,B\in\{+1,-1\}$, in which Alice chooses between two measurement bases $a_1$ and $a_2$, and likewise Bob chooses between the two measurement bases $b_1$ and $b_2$. For systems in particular states subject to judicious choices of measurement bases, the predictions for correlations among the measurement outcomes $A,B$ under various combinations of settings $a_i, b_j$ differ markedly between quantum mechanics and models that satisfy EPR's assumptions of locality and realism.

Subsequently, entangled-particle states have
been shown to violate Bell's inequality in numerous situations, consistent with the predictions of quantum theory \cite{Clauser1978, Larsson14b,brunner2014}. Yet experiments always require sets of assumptions for their interpretation
\cite{Quine1951, Duhem1954}. In tests of local realism, these assumptions can be seen as loopholes, by which, in principle, it could be argued that local realism has not been completely 
ruled out \cite{Larsson14b,Larsson2014}. Closing the locality loophole \cite{Aspect1982, Weihs1998}, for example, requires that the measurement settings are changed by Alice shortly before the arrival of an entangled particle at her detector, such that no signal could inform Bob
about Alice's measurement setting or outcome before Bob completes a measurement at his own detector (and vice versa). The fair sampling assumption, on the other hand, states
that the measured subset of particles is representative of the complete set. This loophole is closed if a sufficiently high fraction of the entangled pairs is detected \cite{Rowe01, Giustina2013, Christensen2013}. Recently, several experiments have observed significant violations of Bell's inequality while simultaneously closing both the locality and fair-sampling loopholes \cite{Hensen15, Giustina2015, Shalm2015, Rosenfeld17}. 

Arguably the most interesting assumption is that the choice of measurement settings is ``free and random," and independent of any physical process that could affect the measurement outcomes \cite{BellSpeakable,bell76,shimony76,bell77,brans88}. 
As Bell himself noted, his inequality was derived under the assumption ``that the settings of instruments are in some sense free variables---say at the whim of experimenters---or in any case not determined in the overlap of the backward light cones" \cite{bell76}. In recent years, this
``freedom-of-choice" loophole has garnered significant theoretical interest \cite{kofler06,hall10,hall11,barrett11,banik12,Gallicchio2014,putz14,putz16,hall16,pironio15}, as well as growing experimental attention \cite{Scheidl2010,aktas15,handsteiner2017a,Wu2016,leung2017,BigBellTest2}.

The freedom-of-choice loophole, as usually understood, concerns events that might have transpired within the causal past of a given experiment, which a local-realist mechanism could have exploited in order to mimic the predictions from quantum mechanics \cite{retrocausalnote}. In a recent pilot test \cite{handsteiner2017a}, measurement settings for a test of Bell's inequality were determined by real-time observation of light from Milky Way stars, thereby constraining any such local-realist mechanism to have acted no more recently than $\sim$\num{600} years ago, rather than microseconds before a given experimental run (as in previous tests \cite{Scheidl2010}). The magnitude of that leap reflected how comparatively little attention had been devoted previously to experimentally addressing this loophole. Given the expansion history of the universe since the big bang, however, the pilot test \cite{handsteiner2017a} excluded only about one hundred-thousandth of one percent of the relevant space-time volume within the past light cone of the experiment.

In this paper, we describe a Cosmic Bell experiment that pushes the origin of the measurement settings considerably deeper into cosmic history, constraining any local-realist mechanism to have acted no more recently than $7.78$ Gyr ago. Based on the arrangement of high-redshift quasars used in our experiment, these results exclude any local-realist mechanism that might have exploited the freedom-of-choice loophole from $96.0\%$ of the space-time volume of the past light cone of the experiment, extending from the big bang to today.

\paragraph{Experimental implementation.}

    \begin{figure*}
	\includegraphics[trim={0cm 0.7cm 0cm 0.4cm}, clip,width=\textwidth]{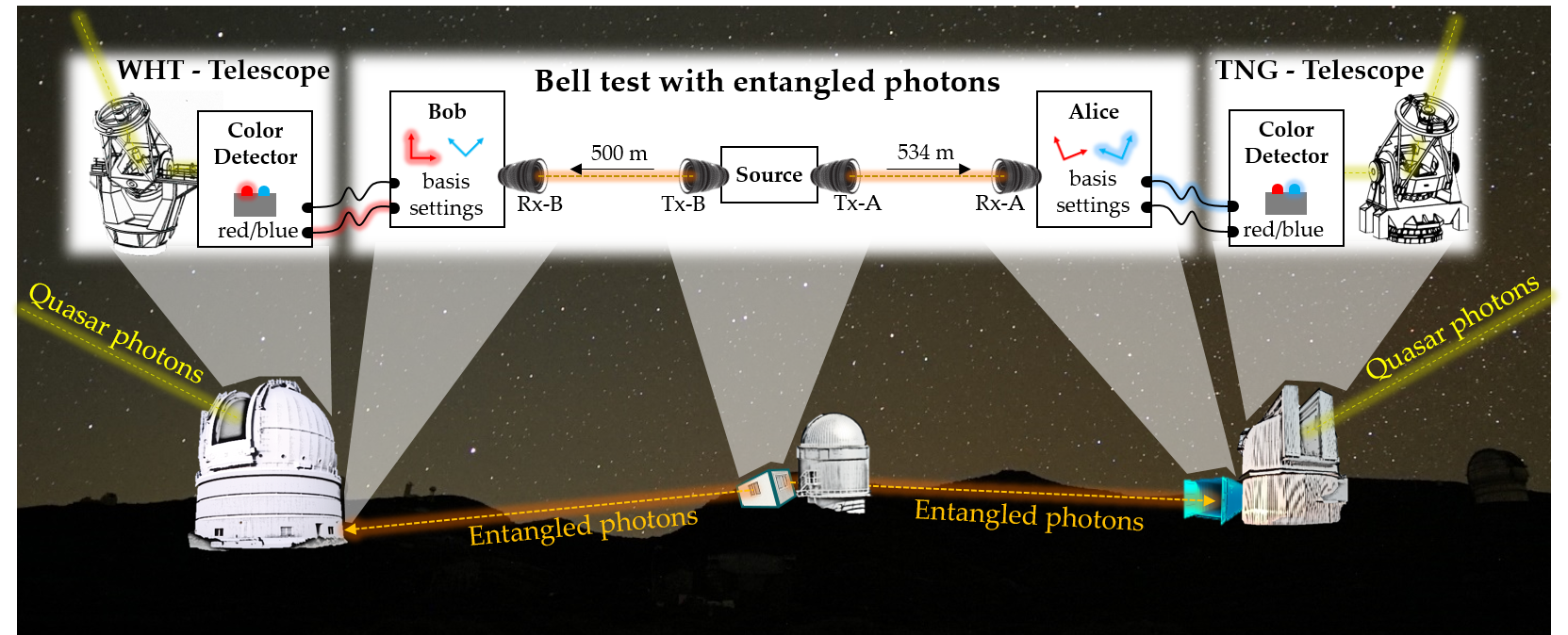}
	\caption{\label{figure1}
	The experimental stations for our Cosmic Bell test. Alice's station received cosmic photons with the \tng (TNG), whose primary mirror diameter is \SI{3.58}{\metre}, while Bob's station received cosmic photons with the \wht (WHT), whose primary mirror diameter is \SI{4.20}{\metre}. Polarization entangled photons were sent from the source to Alice and Bob. Diameters and focal lengths of the quantum channel telescopes were Tx: $d = \SI{70}{\milli\metre}$, $f = \SI{280}{\milli\metre}$; Rx: $d = \SI{140}{\milli\metre}$, $f_{\rm eff} = \SI{1470}{\milli\metre}$. Latitude, longitude, and elevation for the experimental sites were Alice (${\cal A}$: $28.75410^\circ$, $-17.88915^\circ$, $\SI{2375}{\metre}$), Bob (${\cal B}$: $28.760636^\circ$, $-17.8816861^\circ$, $\SI{2352}{\metre}$), and the Source (${\cal S}$: $28.757189^\circ$, $-17.884961^\circ$, $\SI{2385}{\metre}$). The distances from Source to Bob and Alice were \SI{500}{\metre} and \SI{534}{\metre}, respectively. 
	}
	\end{figure*}

\begin{figure*}
	\includegraphics[trim={0cm 0cm 0cm 0cm}, clip,width=\textwidth]{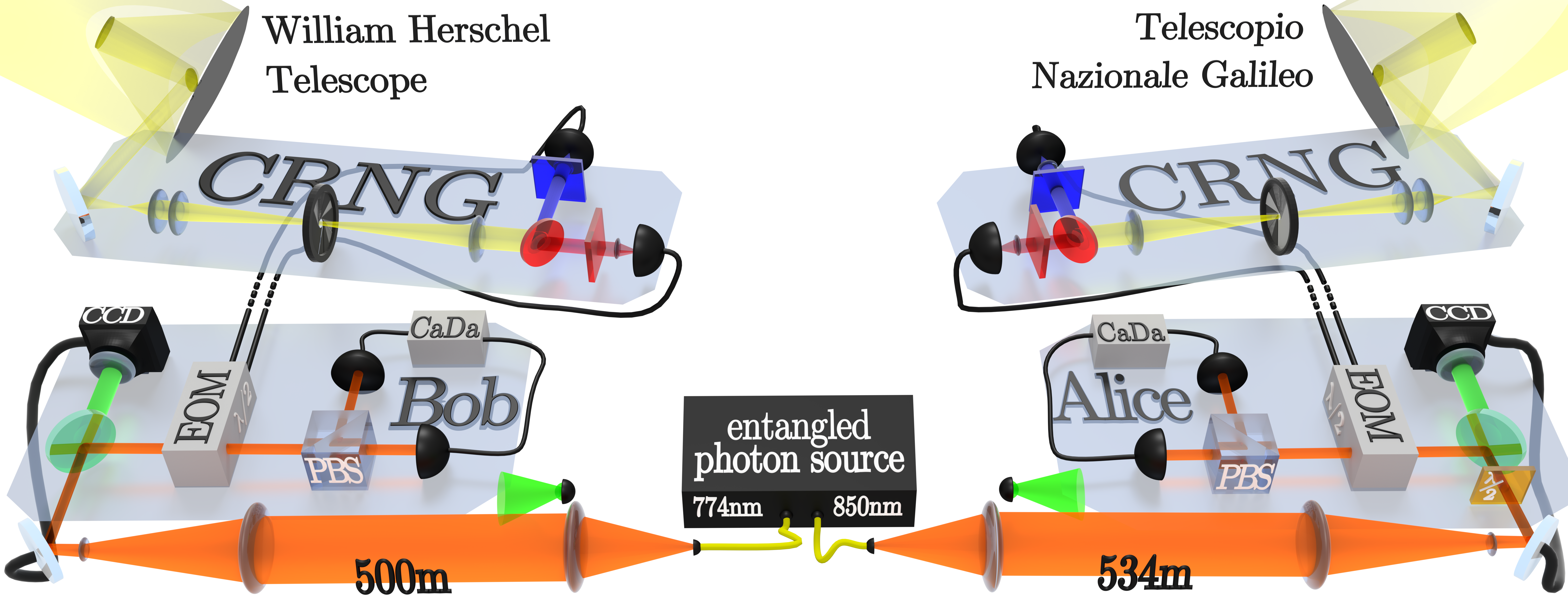}
	\caption{\label{figure2}
	A photon pair source located in the middle produced polarization-entangled photons at center wavelengths of \SI{773.6}{\nano\metre} and \SI{850}{\nano\metre}. The photons were separated into two spatial modes via a dichroic mirror and sent via free-space channels to the quantum receivers at Bob (\SI{773.6}{\nano\metre}) and Alice (\SI{850}{\nano\metre}). Fast steering mirrors guided the photons to the receivers using a green LED as a reference. Electro-optical modulators (EOM) rotate the measurement basis according to the input signals from the CRNGs. Polarization measurements are performed using a polarizing beam splitter
    (PBS) with avalanche photodiodes in each output path. Detection events are timestamped by the control and data acquisition unit (CaDa) and stored locally. Quasar light is collected by the astronomical telescopes and fed into an optical system that creates a magnified image with an iris to restrict the field of view. The quasar light is then split according to its wavelength into a ``blue" and a ``red" channel, whereby each channel contains additional filters to remove misdirected photons. The detector signals are used to trigger the implementation of the corresponding measurement basis at the EOM. 
	}
	\end{figure*}

\Cref{figure1} shows a schematic of the experimental setup at the \orm on the Canary Island of La Palma. A central entangled photon source 
was located in a container next to the \NOT. One entangled-photon observer, Alice, was situated in another container next to the \tng (TNG), and Bob was stationed at the ground floor of the \wht (WHT). The quasar photons were collected by the TNG \cite{tng} and the WHT \cite{wht}. The random numbers extracted from these signals were transmitted to the observers using
%in
BNC cables. The polarization-entangled photons were distributed from the source to the receivers via free-space optical links. A more detailed schematic of the setup can be seen in \Cref{figure2}.

The entangled photon source (see Supplementary Materials \cite{supp}) was based on type-0 spontaneous parametric down-conversion (SPDC) in a Sagnac loop configuration \cite{Steinlechner2014, Kim2006b}. It generated fiber-coupled photon pairs at center wavelengths $\lambda_{\cal A} = \SI{850}{\nano\metre}$ and $\lambda_{\cal B} = \SI{773.6}{\nano\metre}$ in a state close to the maximally entangled Bell state $\ket{\Phi^\pm} = (1/\sqrt{2})(\ket{H_{\cal A} H_{\cal B}}\pm \ket{V_{\cal A} V_{\cal B}})$, where subscripts \alice\ and \bob\ label the respective single-mode fiber for Alice and Bob. 
Each photon was guided to a transmitting telescope (Tx), and distributed via free-space optical channels to the receiving stations of Alice and Bob. Each station consisted of a receiving telescope for entangled photons (Rx), a polarization analyzer (POL), a control and data acquisition unit (CaDa) locked to a rubidium frequency standard, and a cosmic random number generator (CRNG). The entangled photons were guided from the Rx to the polarization analyzer, where an electro-optical modulator (EOM) performed the fast switching between the complementary measurement bases accordingly. The polarization was measured using a polarizing beam splitter followed by a single-photon avalanche diode (SPAD) in each output port.

The CRNGs at TNG (Alice) and WHT (Bob) were essentially identical. The optical path for each CRNG featured a magnified intermediate image, which enabled one to adjust the field of view with an iris in order to minimize background light. 
A dichroic mirror with a cutoff wavelength of \SI{630}{\nano\metre} split the incoming light in a transmitted ``red" and a reflected ``blue" arm. Additional filters (shortpass at \SI{620}{\nano\metre} in the blue arm and longpass at \SI{637}{\nano\metre} in the red arm) efficiently filtered out misdirected photons whose wavelengths were near the cutoff wavelength of the dichroic mirror. Incorporating these additional filters yielded much smaller fractions of misdirected astronomical photons than in our previous experiment \cite{handsteiner2017a}, with $f_{ w} < 2 \times 10^{-5}$ (see the Supplemental Material \cite{supp}). Light from each arm was fed to a SPAD. Electric signals from the SPADs were processed by the CaDa, which triggered the EOM to apply the corresponding measurement settings. Alice measured linear polarization along $22.5^{\circ}/112.5^{\circ}$ (red) and $67.5^{\circ}/157.5^{\circ}$ (blue), while Bob measured linear polarization along $0^{\circ}/90^{\circ}$ (red) and $45^{\circ}/135^{\circ}$ (blue). All SPAD events, from the CRNGs and the polarization analyzers, were timestamped in the CaDa and recorded by a computer. 
	
Using the wavelength of cosmic photons to implement the measurement settings requires the assumption that the wavelength of each photon was set at emission and has not been selectively altered or previewed between emission and detection. (Well-known processes, such as cosmological redshift and gravitational lensing, treat all photons from a given astronomical source in a uniform way, independent of the photons' wavelength at emission \cite{PeeblesBook93,WeinbergBook08,BlandfordNarayan92}.) 

Within an optically linear medium, there does not exist any known physical process that can absorb and re-radiate a given photon at a different wavelength along the same
line of sight, without violating the local conservation of energy and momentum. We further assume that the detected cosmic photons represent a fair sample, despite significant losses in the intergalactic and interstellar media, the Earth's atmosphere, and the experimental setup.

Various scenarios could (in principle) lead to corrupt choices of measurement settings within our experiment. For example, local sources of photons (``noise") rather than genuine astronomical photons could trigger the CRNGs. The most significant sources of local noise include sky glow, light pollution, and detector dark counts. The overall background was measured by pointing each telescope to a dark sky patch \SI{10} arcseconds away from its target quasar before and after each observing period.

\paragraph{Space-time arrangement.} Ensuring locality requires
that any information leaving Alice's quasar at the speed of light along with her setting-determining cosmic photon could not have reached Bob before his measurement of the entangled photon is completed, and vice versa. 

\begin{figure}
	\includegraphics[trim={0cm 0.5cm 0cm 0.8cm}, clip,width = \columnwidth]{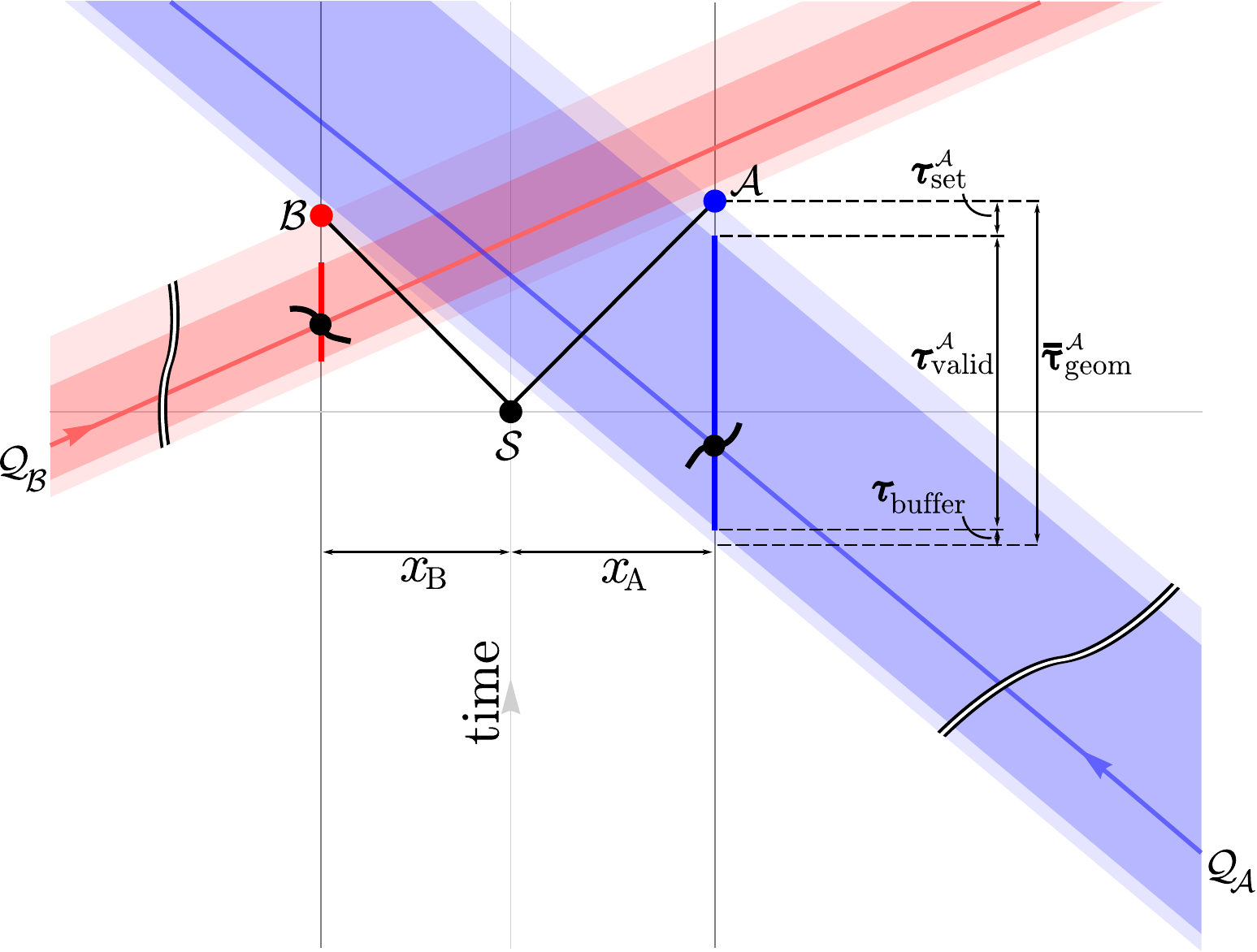}
	\caption{\label{space-time2D} (1+1)D space-time diagram for 
	pair 1, with the origin at the source ${\cal S}$ of entangled pair creation (black dot) and a spatial projection axis chosen to minimize its distance to Alice and Bob. After a short fiber delay (too small to see), entangled photons are sent via free-space channels (black lines) to be measured by Alice and Bob at events $A$ and $B$. Galaxy symbols indicate examples of measurements of valid settings from quasar photons emitted far away at space-time events ${\cal Q}_A$ and ${\cal Q}_B$. Ensuring locality limits settings to the shaded regions. Delays to implement each setting and an added safety buffer shorten the validity time windows that were actually used to the darker shaded regions.}	
\end{figure}

The projected space-time diagram in \Cref{space-time2D} illustrates the situation for the first observed quasar pair (pair 1) of our experiment. The entangled photons are generated at point \source\  and travel through \SI{12}{\metre} of optical fiber, resulting in a delay of $\tau_{\rm fiber}\approx \SI{50}{\nano\second}$.
The distance over the free-space channels is $x_\mathcal{A} = \SI{534}{\metre} $ to \alice\ and $x_\mathcal{B} = \SI{500}{\metre}$ to \bob.

\begin{figure}
	\includegraphics[width = \columnwidth]
	{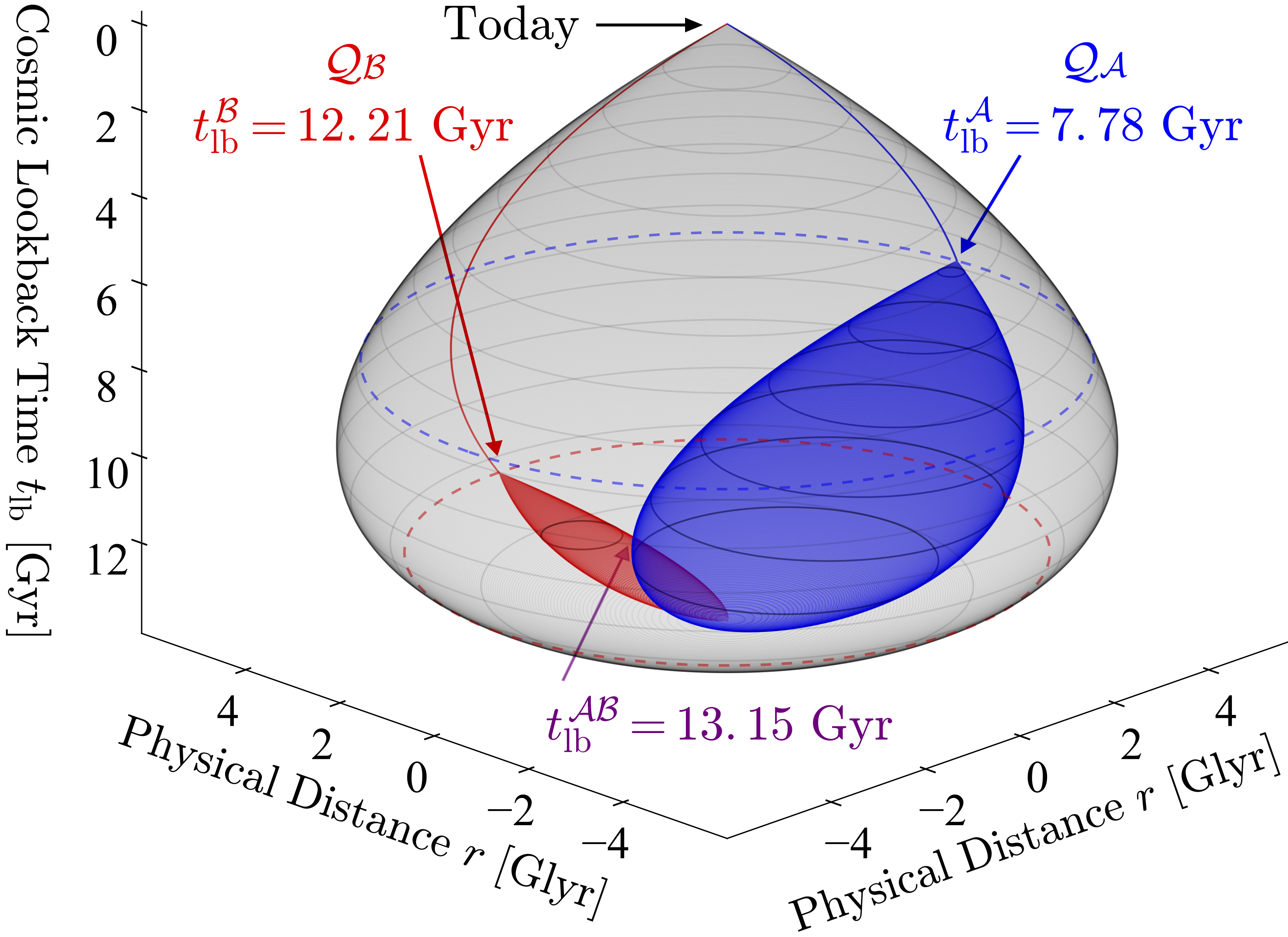}
	\caption{\label{drop} (2+1)D space-time diagram for pair 1, showing the past light cone of our experiment (gray) and of the quasar emission events 
	${\cal Q}_{\cal{A}}$ (blue) and ${\cal Q}_{\cal{B}}$ (red),
	extending back to the big bang, \SI{13.80}{G\yr} ago. 
	The quasars in pair 1 emitted the light that we observed during our experiment $t_{\rm lb}^{\cal A} = \SI{7.78}{G\yr}$ and $t_{\rm lb}^{\cal B} = \SI{12.21}{G\yr}$ ago. The past light cones from ${\cal Q}_{\cal{A}}$ and ${\cal Q}_{\cal{B}}$ last intersected $t_{\rm lb}^{\cal AB} = \SI{13.15}{G\yr}$ ago. The shapes of the light cones reflect the changing rate of cosmic expansion since the big bang.
	To be consistent with our data, any local-realist mechanism would need to have affected detector settings and measurement outcomes of our experiment from within the past light cones of events ${\cal Q}_{\cal A}$, ${\cal Q}_{\cal B}$, or their overlap, a space-time region that consists of only $4.0\%$ of the physical space-time volume contained within the past light cone of our experiment. Such a local-realist scenario would need to have been set in motion at least $7.78$ Gyr ago.
	}
\end{figure}

\begin{table*}
	\caption{\label{quasars} For Alice and Bob's side ($k = \{\mathcal{A}, \mathcal{B} \}$), we list the QSO Simbad identifiers, azimuth (${\rm az}_k$) (clockwise from due North) and altitude (${\rm alt}_k$) above horizon at the start of the observing periods, and redshift ($z$) and lookback time to emission $(t_{\rm lb})$ for quasars observed in pairs 1 and 2, beginning at 
	UTC 2018-01-11 00:20:00 (pair 1) and 2018-01-11 01:21:00 (pair 2). Pair 1 was observed for a total of \SI{17}{\minute}, pair 2 for \SI{12}{\minute}. $\tau^k_{\rm valid}$ is the time the detector setting was valid, taking into account delays and safety margins (see \Cref{space-time2D}).
	The last three columns show the measured CHSH parameter, as well as the $p$ value and the number of standard deviations $\nu$ by which our local-realist model can be rejected (see the Supplemental Material \cite{supp}). 
	}
	\begin{ruledtabular}
		\begin{tabular}{c c l r r c r c c c c}		
			Pair & Side & ID & $\rm az_k^\circ$ & $\rm alt_k^\circ$ & $z$ & $t_{\rm lb}$ [Gyr] & $\tau^k_{\rm valid}$ [\si{\micro\second}] & $S_{\rm exp}$ & $p$-value & $\nu$\\\hline
			\multirow{2}{*}{1} & \alice & QSO B0350-073  & 233 & 38 & \num{0.964} & \num{7.78}  & \num{2.34} & \multirow{2}{*}{2.65} & \multirow{2}{*}{\num{7.4d-21}} & \multirow{2}{*}{\num{9.3}}\\
							   & \bob   & QSO J0831+5245 & 35  & 57 & \num{3.911} & \num{12.21} & \num{0.90} & & & \\
			\multirow{2}{*}{2} & \alice & QSO B0422+004	 & 246 & 38 & \num{0.268} & \num{3.22}  & \num{2.20} & \multirow{2}{*}{2.63} & \multirow{2}{*}{\num{7.0d-13}} & \multirow{2}{*}{\num{7.1}}\\
							   & \bob   & QSO J0831+5245 & 21  & 64 & \num{3.911} & \num{12.21} & \num{0.53} & & & \\
		\end{tabular}
	\end{ruledtabular}
\end{table*}

To ensure the locality conditions, measurements of entangled photons must only be accepted within a certain valid time window, $\tau^k_{\rm valid}$, which has to be chosen such that the selection and implementation of the corresponding settings on one side remain space-like separated from the measurements on the other side.
$\tau^k_{\rm valid}$ is constrained to within a certain time window $\tau^k_{\rm geom}$, which depends on the time-dependent directions of the quasars relative to \alice\ and \bob.  Given the moderate time dependence of $\tau^k_{\rm geom}$ over the relatively brief observing periods ($\leq \SI{17}{\minute}$), we use the shortest value per side within the observing period: $\bar\tau^k_{\rm geom} = \min_t(\tau^k_{\rm geom})$, where $\bar\tau^\mathcal{A}_{\rm geom} = \SI{2.81}{\micro\second}\ (\SI{2.67}{\micro\second})$ and $\bar\tau^\mathcal{B}_{\rm geom} = \SI{1.48}{\micro\second}\ (\SI{1.11}{\micro\second})$ for pair 1 (2). Various delays from signal transmission through fibers
and BNC cables, and to implement a given setting with the EOM, have to be 
subtracted from $\tau^k_{\rm geom}$ to compute the correct validity time $\tau^k_{\rm valid}$.
The delay until a certain setting was implemented, $\tau_{\rm set}^k$, was measured to be \SI{325}{\nano\second} and \SI{430}{\nano\second} for Alice and Bob, respectively.
An additional buffer was used on both sides with $\tau_{\rm buffer} = \SI{150}{\nano\second}$ to account for small inaccuracies in timing and distance measurements (for details please refer to the Supplemental Material \cite{supp}). The final validity time we used is then $\tau^k_{\rm valid} = \bar{\tau}^k_{\rm geom} - \tau^k_{\rm set} - \tau_{\rm buffer}$.

For pairs 1 and 2, measurement settings at Bob's station were determined based on observations of a quasar with redshift $z_{\cal B} = 3.911$ \cite{Downes99}, corresponding to a lookback time to the emission of that light $t_{\rm lb}^{\cal B} = \SI{12.21}{G\yr}$ ago. Measurement settings at Alice's station were determined based on observations of quasars with $z_{\cal{A}} = 0.964$ \cite{Paris17} (pair 1) and $z_{\cal{A}} = 0.268$ \cite{Shaw13} (pair 2), corresponding to $t_{\rm lb}^{\cal A} = \SI{7.78}{G\yr}$ and \SI{3.22}{G\yr} ago, respectively. (See \Cref{quasars}.) These times may be compared with the age of our observable universe since the big bang, $t_{\rm lb} = \SI{13.80}{G\yr}$ \cite{ade2016planck}. We consider possible implications of inhomogeneities along the lines of sight to these objects, such as gravitational lensing effects, in the Supplemental Material \cite{supp}.

\Cref{drop} depicts the past light cone of our experiment (gray) together with the past light cones of quasar emission events ${\cal Q}_{\cal{A}}$ (blue) and ${\cal Q}_{\cal{B}}$ (red) for the quasars of pair 1. The past light cones from ${\cal Q}_{\cal{A}}$ and ${\cal Q}_{\cal{B}}$ for this pair last intersected $t_{\rm lb}^{\cal AB} = \SI{13.15}{G\yr}$ ago, less than $650$ million years after the big bang. (For pair 2, the past light cones most recently intersected $t_{\rm lb}^{\cal AB} = \SI{12.47}{G\yr}$ ago.) This is the most recent time by which a correlation between the two quasars could have occurred or been orchestrated. The space-time 4-volume contained within the union of the past light cones from ${\cal Q}_{\cal{A}}$ and ${\cal Q}_{\cal{B}}$ constitutes just 
$4.0\%$ (pair 1) and $36.5\%$ (pair 2) 
of the 4-volume within the past light cone of our experiment. (See Supplemental Material \cite{supp}.)
Events associated with any local-realist mechanism that could have affected detector settings and measurement outcomes of our experiment
would need to lie within the past light cones of ${\cal Q}_{\cal{A}}$ and/or ${\cal Q}_{\cal{B}}$,
and hence are restricted to have acted no more recently than $t_{\rm lb}^{\cal A} = \SI{7.78}{G\yr}$ or \SI{3.22}{G\yr} ago for pairs 1 and 2, respectively.

\paragraph{Analysis and results.} We performed two Cosmic Bell tests with the quasars listed in \Cref{quasars}, for a total measurement time of \SI{17}{\minute} (pair 1) and \SI{12}{\minute} (pair 2). 
In the analysis of our acquired data, we follow the assumption of fair sampling and fair coincidences \cite{Larsson2014}. Thus, our data can be postselected for coincidence events at Alice's and Bob's stations. We correct for the clock drift as in Ref.~\cite{Scheidl2009} and identify coincidences within a time window of \SI{2.66}{\nano\second}. We then check for correlations between measurement outcomes $A,B\in \{+1, -1\}$ for particular settings choices $(a_i,b_j), i,j \in \{1,2\}$ using the Clauser-Horne-Shimony-Holt (CHSH) inequality \cite{Clauser1969}: 
\begin{equation}
	S\equiv|E_{11} + E_{12} + E_{21} - E_{22}| \leq 2,
\end{equation}
where $E_{ij} = 2p(A = B|a_ib_j)-1$ and $p(A=B|a_ib_j)$ is the probability of Alice and Bob obtaining the same measurement outcome for the joint settings $(a_i,b_j)$. While four probabilities can arithmetically add up to \num{4}, local-realistic correlations cannot exceed an $S$ value of $2$ and the quantum-mechanical limit is $2\sqrt{2}$ \cite{Cirelson80}. 

As can be seen from \Cref{quasars}, the measured $S_{\rm exp}$ values are \num{2.65} and \num{2.63} for pairs 1 and 2, which clearly
exceed the local-realist bound of $2$. However, not all of our settings were determined by genuine cosmic photons. A certain fraction of settings $\epsilon_k$ on each side ($k\in\{\mathcal{A},\mathcal{B}\}$) was produced by some kind of local process, including sky glow, ambient light, and detector dark counts. 
We therefore consider such settings to be ``corrupt" and assume that a local-realist mechanism could have exploited them to produce maximal CHSH correlations, with $S = 4$. Such a (hypothetical) mechanism could produce CHSH correlations as large as $S = 2 (1 - \epsilon_{\cal A} - \epsilon_{\cal B}) + 4 (\epsilon_{\cal A} + \epsilon_{\cal B})$ \cite{Kofler16,handsteiner2017a,leung2017}.  

In our analysis we account for such ``corrupt" settings as well as unequal (biased) frequencies for various combinations of detector settings $(a_i, b_j)$, and possible ``memory effects" by which a local-realist mechanism could exploit knowledge of settings and outcomes of previous trials (see Supplemental Material \cite{supp}).
From this detailed treatment, we find that correlations at least as large as observed in our data
could have been produced by a local-realist mechanism only with probabilities $p\leq\num{7.4d-21}$ for pair 1 and $p\leq\num{7.0d-13}$ for pair 2, corresponding to experimental violations of the Bell-CHSH bound by at least \num{9.3} and \num{7.1} standard deviations, respectively.

\paragraph{Conclusions.} For each Cosmic Bell test reported here, we assume fair sampling and close the locality loophole. We also constrain the freedom-of-choice loophole with detector settings determined by extragalactic events, such that any local-realist mechanism would need to have acted no more recently than \SI{7.78}{\giga\yr} or \SI{3.22}{\giga\yr} ago for pairs 1 and 2, respectively---more than six orders of magnitude deeper into cosmic history than the experiments reported in Ref.~\cite{handsteiner2017a}. This corresponds to excluding such local-realist mechanisms from $96.0\%$ (pair 1) and $63.5\%$ (pair 2) of the relevant space-time regions, compared to $\sim 10^{-5}\%$ of the relevant space-time region as in Ref.~\cite{handsteiner2017a} (see Supplemental Materials \cite{supp}).

We have therefore dramatically limited the space-time regions from which local-realist mechanisms could have affected the outcome of our experiment
to early in the history of our universe. To constrain such models further, one could use other physical signals to set detector settings, such as patches of the cosmic microwave background radiation (CMB), or even primordial neutrinos or gravitational waves, thereby constraining such models all the way back to the big bang---or perhaps even earlier, into a phase of early-universe inflation \cite{Gallicchio2014,handsteiner2017a}. Such extreme tests might ultimately prove relevant to the question of whether quantum entanglement undergirds the emergence of space-time itself. (For a recent review, see Ref.~\cite{VanRaamsdonk:2016exw}).

\paragraph{Note Added.} After we completed our experiment, a similar experiment was conducted by another group, the results of which are reported in Ref.~\cite{LiPan18}.
	
\paragraph{Acknowledgements.} The authors would like to thank Cecilia Fari\~na, \'{E}milie L'Hom\'{e}, Karl Kolle, Neil O'Mahony, J\"urg Rey, Fiona Riddick and the whole team at the WHT as well as Emilio Molinari, Giovanni Mainella, Carlos Gonzalez and the whole team at the TNG for their tremendous support of our experiment. We also thank Thomas Augusteijn, Carlos Perez and all the staff at the NOT for their support, which did not decrease even after our container crashed into their telescope in a storm. We are also grateful for the encouraging support of Cesare Barbieri. In addition, we are grateful to Brian Keating, Hien Nguyen, Paul Schechter, and Gary Cole for helpful discussions. This work was supported by the Austrian Academy of Sciences (OEAW), by the Austrian Science Fund (FWF) with SFB F40 (FOQUS) and FWF project CoQuS No.~W1210-N16, the Austrian Federal Ministry of Education, Science and Research (BMBWF) and the University of Vienna via the project QUESS. This work was also supported by NSF INSPIRE Grant no.~PHY-1541160. Portions of this work were conducted in MIT's Center for Theoretical Physics and supported in part by the U.S. Department of Energy under Contract No.~DE-SC0012567. C.L. was supported by the U.S. Department of Defense (DoD) through the National Defense Science \& Engineering Graduate Fellowship (NDSEG) Program.

\clearpage

\onecolumngrid
\appendix

\setcounter{equation}{0}
\setcounter{table}{0}
\setcounter{figure}{0}

\centerline{\large\bf Supplemental Material }
\vspace{0.15cm}
\centerline{\large\bf Cosmic Bell Test using Random Measurement Settings from High-Redshift Quasars}
\vspace{0.5cm}

\twocolumngrid

\section{Causal Alignment}

As described in Ref.~\cite{handsteiner2017a}, the time-dependent locations of astronomical sources on the sky relative to our ground-based experimental site complicates the enforcement of the space-like separation conditions needed to address both the locality and freedom-of-choice loopholes. For example, the photon from quasar emission event ${\cal Q}_A$ must be received by Alice's cosmic-photon receiving telescope (Rx-CP) before that photon's causal wavefront reaches either the Rx-CP or the entangled-photon receiving telescope (Rx-EP) on Bob's side, and vice versa.

In this section we first present our main result for the causal-alignment windows, $\tau^k_{\rm valid} (t)$ (for sides $k, \ell = \{ {\cal A, B } \}$), within which settings chosen by astronomical photons remain valid, and then derive the various terms in our expression. As shown in Figure 3 of the Main paper, we parameterize $\tau_{\rm valid}^k (t)$ as
%%%%%%
\begin{equation}
    \tau_{\rm valid}^k (t) = \bar{\tau}_{\rm geom}^k (t) - \tau_{\rm set}^k - \tau_{\rm buffer}^k ,
    \label{tauvaliddef}
\end{equation}
where $\tau_{\rm geom}^k (t)$ arises from the geometrical arrangement of the quasars relative to the locations of relevant instrumentation on Earth, and $\bar{\tau}_{\rm geom}^k (t) \equiv {\rm min}_t [\tau_{\rm geom}^k (t)]$ is the minimum value of $\tau_{\rm geom}^k (t)$ during an observing window. The term $\tau_{\rm set}^k$ indicates the time required to electronically output a bit and implement the detector setting, while $\tau_{\rm buffer}^k$ accommodates total delays due to atmosphere, telescope optics, and detector response. Negative validity times for either $k$ would indicate an instantaneous configuration that was out of ``causal alignment," in which at least one setting would be invalid for the purposes of closing the locality loophole.

For $\tau_{\rm geom}^k (t)$, we find
\begin{eqnarray}
%\begin{split}
\tau^k_{\rm geom} (t) &=& \frac{1}{c} \hat{\bf n}_k (t) \cdot ( {\bf r}_k - {\bf m}_\ell ) + \frac{ n}{c} \bigg[ \vert {\bf m}_k - {\bf s} \vert - \vert {\bf m}_\ell - {\bf s} \vert \bigg] \nonumber \\
& \quad \quad& - \frac{ \gamma_k}{c} \vert {\bf r}_k - {\bf m}_k \vert ,
\label{taukfinal}
\end{eqnarray}
where ${\bf r}_k$ and ${\bf m}_k$ are the spatial 3-vectors for the locations of the cosmic receiving telescopes (Rx-CP) and entangled-particle detectors (Rx-EP) 
for side $k$, respectively; ${\bf s}$ is the spatial 3-vector for the location of the source of entangled particles; and $c$ is the speed of light in vacuum. The time-dependent unit vector $\hat{\bf n}_k (t)$ points toward the quasar used to set detector settings on side $k$, and is computed using astronomical ephemeris calculations. Additionally, $n$ is the index of refraction of air and $\gamma_k$, which acts like an index of refraction, parameterizes  the group velocity delay through fiber optics and/or electrical cables connecting the telescope and entangled photon detector on side $k$.

We work with a space-time metric signature $(+, -, -, -)$, so that space-time events represented by four-vectors ${\cal A}^\mu$ and ${\cal B}^\mu$ will be space-like separated if $({\cal A}^\mu - {\cal B}^\mu)^2 < 0$. We represent spatial and temporal intervals of cosmological magnitude---such as the interval between emission of light from a distant quasar and its detection on Earth today---in terms of a spatially flat Friedmann-Lema\^{i}tre-Robertson-Walker (FLRW) line element because on length-scales greater than ${\cal O} (100)$ Mpc, our universe has been measured to be homogeneous \cite{Sarkar09}, isotropic \cite{Marinoni12}, and spatially flat \cite{ade2016planck} to high accuracy. We have
%%%%%
\begin{eqnarray}
ds^2 &=& g_{\mu\nu} dx^\mu dx^\nu  \nonumber \\
&=& c^2 dt^2 - R_0^2 a^2 (t) \left[ d\chi^2 + \chi^2 d\Omega_{(2)}^2 (\theta, \phi) \right] ,
\label{ds1}
\end{eqnarray}
where $t$ is cosmic time, equal to the proper time recorded by a freely falling observer at the origin of the spatial coordinate system, $\chi$ is a (dimensionless) comoving distance, and $d\Omega_{(2)}^2 (\theta, \phi) = d\theta^2 + \sin^2 \theta d \phi^2$ is the line-element for a unit $2$-sphere. In this section we ignore possible complications from inhomogeneities, such as gravitational-lensing effects.

In Eq.~(\ref{ds1}), $a(t)$ is the (dimensionless) cosmic scale factor and the constant $R_0 \equiv c / H_0$ has dimensions of length. We use $H_0 = 67.74 \, {\rm km} \, {\rm s}^{-1} \, {\rm Mpc}^{-1}$ as the present value of the Hubble parameter \cite{ade2016planck}, corresponding to a Hubble time $t_H = H_0^{-1} = 14.43$ Gyr and hence $R_0 = 14.43$ Glyr. Physical distances at a given cosmic time, $r (t)$, are related to comoving distances by $r (t) = R_0 a (t) \chi$. The observed redshift, $z$, for astronomical objects arising from cosmic expansion is given by
%%%%%%%%
\begin{equation}
1 + z = \frac{ a(t_0) }{a (t_e) } ,
\label{zdef}
\end{equation}
where $t_0$ is the present time and $t_e$ is the time of emission. We set $a (t_0) = 1$ and take $t = 0$ to be the time of the hot big bang (following any primordial phase of inflation, if inflation occurred). We further assign the origin of the spatial coordinates to be the center of the Earth. Errors introduced by treating the rotating Earth as an inertial frame are less than one part in $10^6$, and are easily accommodated within $\tau_{\rm buffer}$.

\begin{figure}
\centering
\includegraphics[width=0.49\textwidth]{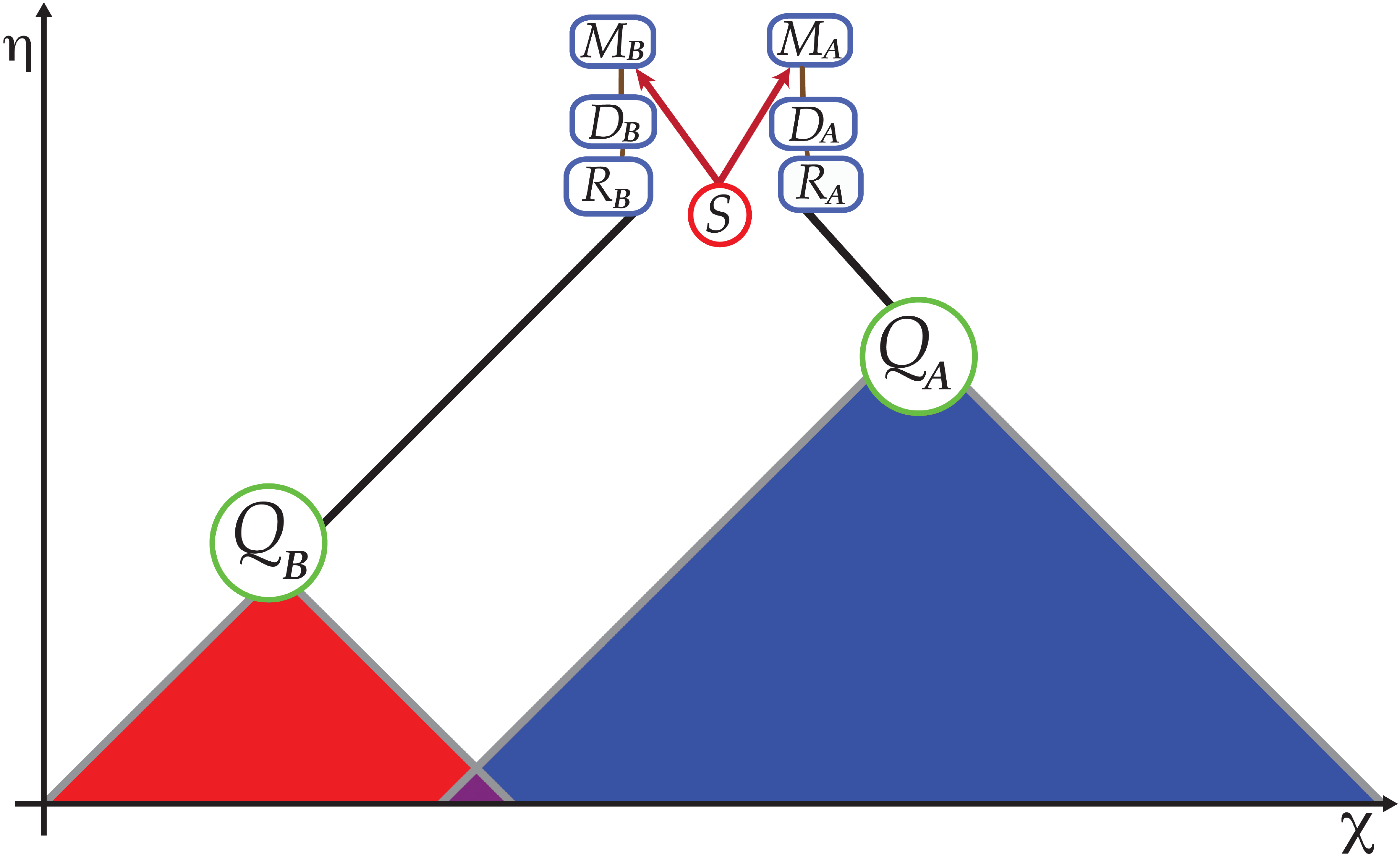}
\caption{\small 
Schematic space-time diagram (not to scale) of our cosmic Bell test, in (dimensionless) conformal time $\eta$ versus comoving distance $\chi$. In these coordinates, null geodesics appear as $45^\circ$ diagonals. On each side $k \in  \{ A, B\}$, light from quasar emission event ${\cal Q}_k$ is received on Earth at event ${\cal R}_k$ and used to determine the detector setting at event ${\cal D}_k$. Meanwhile, spacelike-separated from events ${\cal Q}_k$, ${\cal R}_k$, and ${\cal D}_k$, the source ${\cal S}$ emits a pair of entangled particles, which are measured at events ${\cal M}_k$. 
}
\label{fig:conformal1}
\end{figure}

Quasar emission event ${\cal Q}_k$ occurred a long time ago, in a galaxy far, far away. Hence it is convenient to introduce (dimensionless) conformal time, $d\eta \equiv H_0 dt /  a(t) $, to take into account the cosmic expansion between the emission and detection of the cosmic photons. Then Eq.~(\ref{ds1}) becomes
%%%%%%
\begin{equation}
ds^2 = R_0^2 a^2 (\eta) \left[ d\eta^2 - d\chi^2 - \chi^2 d\Omega_{(2)}^2 \right] ,
\label{dseta}
\end{equation}
and (radial) null geodesics correspond to $d \eta = d \chi$. In these coordinates, the 4-vector corresponding to the receipt of a quasar photon at detector $k$ on Earth may be written ${\cal R}_k^\mu = (\eta_{r_k}, {\bm \chi}_{r_k} )$, and the 4-vector for the emission of the photon at the quasar is
%%%%%%
\begin{equation}
{\cal Q}_k^\mu = \left(\eta_{q_k}, {\bm \chi}_{r_k} + \left(\eta_{r_k} - \eta_{q_k} \right) \hat{\bf n}_k \right) ,
\label{Qk}
\end{equation}
where ${\bm \chi}_{r_k}$ is the (comoving) spatial location of the quasar photon detector Rx-CP on side $k$, and $\hat{\bf n}_k (t)$ is the unit spatial 3-vector pointing from the center of the Earth toward the quasar. The entangled pair is emitted from the source at ${\cal S}^\mu = ( \eta_s , {\bm \chi}_s )$. See \Cref{fig:conformal1}.

The entangled-pair emission event should be space-like separated from the arrival of the quasar photons, which requires $( {\cal Q}_k - {\cal S})^2 < 0$.
Assuming $(\eta_{r_k} - \eta_s) , \vert {\bm \chi}_{r_k} - {\bm \chi}_s \vert \ll (\eta_{r_k} - \eta_{q_k} )$, this implies
%%%%%%%
\begin{equation}
- \hat{\bf n}_k \cdot \left( {\bm \chi}_{r_k} - {\bm \chi}_s \right) < \eta_{r_k} - \eta_s .
\label{chiretar1}
\end{equation}
(We further note that in the limit $(\eta_{r_k} - \eta_{q_k}) \gg \vert {\bm \chi}_{r_k} \vert$, the spherical waves emitted from the quasar arrive at the Earth as plane waves to a very good approximation.) The quantities in Eq.~(\ref{chiretar1}) all refer to Earthbound events, and hence we may transform back to coordinates more convenient for describing a given experimental trial. Upon recalling that $a (t_0) = 1$, we have ${\bm \chi}_{r_k} = R_0^{-1} {\bf r}_k$, where ${\bf r}_k$ is the present (physical) spatial location of the quasar-photon detector Rx-CP as reckoned from the center of the Earth, and likewise ${\bm \chi}_s = R_0^{-1} {\bf s}$. We also have
%%%%%%%
\begin{equation}
\eta_{r_k} - \eta_s = H_0 \int_{t_s}^{t_{r_k}} \frac{ dt'}{a (t' ) } .
\label{etarketas}
\end{equation}
The cosmic scale factor varies imperceptibly during the course of the experiment, so we may expand $a (t_0 + \delta t) = a (t_0) + \delta t / t_H + {\cal O} [ (\delta t / t_H )^2 ]$, given that $t_H^{-1} = H_0 = \dot{a} (t_0) / a(t_0)$, and overdots denote derivatives with respect to cosmic time, $t$. During a given experimental trial, $\delta t = t_{r_k} - t_s$ is typically a fraction of a second, so we have $(\eta_{r_k} - \eta_s ) = H_0 (t_{t_k} - t_s ) + {\cal O} (10^{-17})$. Then Eq.~(\ref{chiretar1}) becomes
%%%%%
\begin{equation}
- \frac{ \hat{\bf n}_k}{c}  \cdot ({\bf r}_k - {\bf s} ) <  t_{r_k} - t_s  .
\label{trts1}
\end{equation}

Each measurement on an entangled particle should be completed before the causal wavefront from the quasar emission on the other side arrives, which requires $({\cal Q}_k - {\cal M}_\ell )^2 < 0$. Representing ${\cal M}_k^\mu = (\eta_{m_k} , {\bm \chi}_{m_k})$, using Eq.~(\ref{Qk}) for ${\cal Q}_k^\mu$, and proceeding as above to relate $(\eta_{m_k}, {\bm \chi}_{m_k})$ to $(t_{m_k} , {\bf r}_{m_k})$, we find
%%%%%%%
\begin{equation}
(t_{r_k} - t_{m_\ell} ) + \frac{ \hat{\bf n}_k }{c} \cdot ({\bf r}_k - {\bf m}_\ell ) > 0 .
\label{trtm}
\end{equation}
Meanwhile, each quasar photon must be received, processed, and converted to a stable detector setting before the entangled photon from the source arrives. To calculate $\tau_{\rm geom}^k (t)$, we consider only the spatial arrangement of the various events, since we separately accommodate additional delays (from telescope optics, electronics, cables, and the like) with the factors $\tau_{\rm set}^k$ and $\tau_{\rm buffer}^k$. For $\tau_{\rm geom}^k (t)$, we may therefore parameterize
\begin{equation}
t_{d_k} - t_{r_k} = \frac{ \gamma_k}{c} \vert {\bf r}_k - {\bf m}_k \vert ,
\label{deltak}
\end{equation}
where $t_{d_k}$ is the time when the setting for the entangled-photon detector on side $k$ is set. Eq.~(\ref{deltak}) takes into account the fact that the quasar-photon reception and the detector-setting event can occur at different spatial locations.
For measurements on the future light cone of the entangled-particle emission, we can use the particles' travel time from the source to the detectors to write
%%%%%%%
\begin{equation}
t_{m_k} = t_s + \frac{ n}{c} \vert {\bf m}_k - {\bf s} \vert .
\label{tmk}
\end{equation}
For the setting to be valid, meanwhile, it must be set before the measurement, $t_{d_k} < t_{m_k}$. Then we may rearrange Eqs.~(\ref{deltak}) and (\ref{tmk}) to write
%%%%%%%%
\begin{equation}
t_{r_k} - t_s < \frac{ n}{c} \vert {\bf m}_k - {\bf s} \vert - \frac{ \gamma_k}{c} \vert {\bf r}_k - {\bf m}_k \vert .
\label{trts2}
\end{equation}
Similarly substituting Eq.~(\ref{tmk}) for $t_{m_\ell}$ into Eq.~(\ref{trtm}) we have
\begin{equation}
 \frac{ n}{c} \vert {\bf m}_\ell - {\bf s} \vert - \frac{ \hat{\bf n}_k}{c} \cdot ({\bf r}_k - {\bf m}_\ell ) < t_{r_k} - t_s .
\label{trts3}
\end{equation}
We want the most conservative limit on $\tau_{\rm geom}^k (t)$. Subtracting the lefthand side of Eq.~(\ref{trts1}) from the lefthand side of Eq.~(\ref{trts3}) yields $c^{-1} (n + \cos \theta ) \vert {\bf m}_\ell - {\bf s} \vert \geq 0$, since the index of refraction satisfies $n \geq 1$, and $\vert \cos \theta \vert \leq 1$, where $\theta$ is the angle between $\hat{\bf n}_k$ and $({\bf m}_\ell - {\bf s} )$. Therefore the lefthand side of Eq.~(\ref{trts3}) provides the tighter lower bound on $(t_{r_k} - t_s)$. The validity time is determined by the difference between the upper and lower bounds on $(t_{r_k} - t_s)$: subtracting the lefthand side of Eq.~(\ref{trts3}) from the righthand side of Eq.~(\ref{trts2}) yields our expression for $\tau_{\rm geom}^k (t)$ in Eq.~(\ref{taukfinal}).

For our experiment, 
we may set ${\bf m}_k \simeq {\bf r}_k$ for each side, and accommodate measured delays for signal propagation and processing within the factors $\tau_{\rm set}^k$. Then we may compute values for $\tau_{\rm geom}^k (t)$ using the coordinates for the various experimental stations shown in \Cref{tab:coordinates}. We used $\tau_{\rm buffer} = 150$ ns on each side, as well as $\tau_{\rm set}^{\cal A} = 325$ ns and $\tau_{\rm set}^{\cal B} = 430$ ns. Incorporating these values for $\tau_{\rm set}^k$ and $\tau_{\rm buffer}$ as in Eq.~(\ref{tauvaliddef}) yielded the final values of 
$\tau_{\rm valid}^k$ that we used, shown in \Cref{tab:quasars}.

\begin{table}[ht]
    \centering
    \begin{tabular}{|c|c | c | c | c |}
    \hline
    Component & Lat.$^{\circ}$ & Lon.$^{\circ}$ & Elev. [m] & ${\rm diam}_k$ [m] \\
    \hline
    Rx-CP ${\cal A}$ & $28.75410^{\circ}$ & $-17.88915^{\circ}$ & $2375$ & $3.58$ \\
    Source ${\cal S}$ & $28.757189^{\circ}$ & $-17.884961^{\circ}$ & $2385$ &  \\
    Rx-CP ${\cal B}$ & $28.760636^{\circ}$ & $-17.8816861^{\circ}$ & $2352$ & $4.20$ \\
    \hline
    \end{tabular}
    \caption{\small Latitude, Longitude, and Elevation for the cosmic-photon receiving telescopes (Rx-CP) 
    for Alice (${\cal A}$) and Bob (${\cal B}$), and for the Source ($S$). Also shown are the aperture diameters (${\rm diam}_k$) for the %various 
    telescopes. 
    }
    \label{tab:coordinates}
\end{table}

\begin{table*}[ht]
%\footnotesize
\centering
\begin{tabular}{| c | c  | l | r | r | r | r | r | c | c | c | c | c |  } 
\hline
Pair & Side & Simbad ID  & RA$^\circ$  & DEC$^\circ$  & $\alpha^{\circ}$ & az$_k^\circ$ & alt$_k^\circ$  & $z$ & $\eta_k$ & $t^{k}_{\rm lb}$ (Gyr) &  $F_{\rm excl}$ & $\tau^k_{\rm valid}$ [$\mu$s]  \\
\hline           
$1$ &  ${\cal A}$ & QSO B0350-073 & $58.127300$ & $-7.183976$  & 83.81 & 233 & 38  & 0.964  & 2.46 & 7.78 & 0.960 & 2.34  \\ 
    & ${\cal B}$ & QSO J0831+5245 & $127.923750$ & $52.754860$ &       & 35 & 57 & $3.911$ & $1.56$ & $12.21$ & & $0.90$ \\ 
 \hline
$2$ & ${\cal A}$ & QSO B0422+004 & $66.195175$ & $0.601758$   & 72.84 & 246 & 38 & $0.268$ & $2.95$ & $3.22$ & 0.635 & $2.20$ \\
   & ${\cal B}$ & QSO J0831+5245 & $127.923750$ & $52.754860$ &       & 21  & 64 & $3.911$ & $1.56$ & $12.21$ & & $0.53$ \\ 
\hline
\end{tabular}\par
\caption{
\small
Quasars whose light was used to determine detector settings for Alice (${\cal A}$) and Bob (${\cal B}$) for pairs 1 and 2. For each quasar, we list its QSO ID number from the Simbad database, celestial coordinates, angular separations $(\alpha)$ of each pair, azimuth (clockwise from due North) and altitude above horizon at the start of each observating run, and redshift ($z$). We also list the (dimensionless) conformal time of quasar emission 
($\eta_k$),
the lookback times (in Gyr) to each quasar emission event ($t^{k}_{\rm lb}$), 
as well as the fraction of the physical 4-volume of the past light cone of our experiment, extending back to the big bang, from which any local-realist mechanism that might account for the measured violations of the Bell-CHSH inequality is excluded ($F_{\rm excl})$.
Finally, we list the validity time $\tau^k_{\rm valid}$ from Eq.~(\ref{tauvaliddef}), which gives the minimum time that detector settings are valid for side $k=\{ {\cal A}, {\cal B} \}$ during each experimental run, taking into account various delays and safety margins. 
}
\label{tab:quasars}
\end{table*}

\section{Excluded Spacetime Regions}

By using light from distant quasars to determine detector settings, we may constrain the space-time region within which any putative local-realist mechanism could have engineered the observed correlations among measurements on the entangled particles. In this section we consider the space-time regions excluded from such local-realist scenarios.

Following the discussion in Ref.~\cite{Friedman2013}, we may relate the measured redshift for a given astronomical object to the conformal time at which the light we receive on Earth was emitted by the object, $\eta_q$. We take $\eta = 0$ to correspond to the time of the hot big bang. We may also compute the lookback time to the emission event (in cosmic time), $t_{\rm lb}$, reckoned from the present, $t_0$. 

We parameterize the Friedmann equation governing the evolution of $a(t)$ in terms of the function
%%%%%%
\begin{equation}
E(a) \equiv \frac{ H(a) }{H_0} = \sqrt{ \Omega_\Lambda + \Omega_k a^{-2} + \Omega_M a^{-3} + \Omega_R a^{-4} } ,
\label{Eadef}
\end{equation}
where $H (a)$ is the Hubble parameter for a given scale factor $a = a(t)$, and we again use the best-fit value $H_0 = 67.74 \> {\rm km} \, {\rm s}^{-1} \, {\rm Mpc}^{-1}$ \cite{ade2016planck}. The $\Omega_i \equiv \rho_i / \rho_c$ are the present-day ratios of the energy densities of dark energy ($\rho_\Lambda$), cold matter ($\rho_M$), and radiation ($\rho_R$) to the critical density, $\rho_c = 3 H_0^2 / (8 \pi G)$, where $G$ is Newton's gravitational constant. (The quantity $\rho_M$ includes contributions from both baryonic matter and cold dark matter.) We also define the total fractional density of dark energy, cold matter, and radiation ($\Omega_T \equiv \Omega_\Lambda + \Omega_M + \Omega_R$), and the fractional density associated with spatial curvature ($\Omega_k \equiv 1 - \Omega_T$). We assume that $\rho_\Lambda$ arises from a genuine cosmological constant with equation of state $w = p / \rho = -1$, and hence $\Omega_\Lambda a^{-3 (1 + w) } = \Omega_\Lambda$, which is consistent with observations \cite{ade2016planck}. We adopt the best-fit cosmological parameters from Ref.~\cite{ade2016planck},
%%%%%%%
\begin{equation}
\vec{\Omega} = (\Omega_\Lambda, \Omega_M, \Omega_R) 
= (0.6911,0.3089,9.16 \times 10^{-5} ), 
\label{vecOmega}
\end{equation}
consistent with $\vert \Omega_k \vert < {\cal O} (10^{-3})$. Here $\Omega_R = \Omega_M / (1 + z_{\rm eq})$, with the redshift for matter-radiation equality given by $z_{\rm eq} = 3371$ \cite{ade2016planck}.  

Redshifts for the three quasars we observed are listed in \Cref{tab:quasars}. For QSO J0831+5245, which was observed for both quasar pairs, we use a reported host galaxy redshift of $z=3.9114 \pm 0.0003$ from Ref.~\cite{Downes99}. For QSO B0350-073, we use the reported redshift of $z=0.9635389 \pm 0.00011$ from the Sloan Digital Sky Survey (SDSS) Quasar Catalog Fourteenth Data Release \cite{Paris17}. For QSO B0422+004, reported redshifts include $z=0.31$ \cite{Mingaliev14} and $z=0.268$ \cite{Shaw13}. Neither reported redshift included an uncertainty, so we conservatively adopt the smaller value, $z=0.268$. Given the small redshift uncertainties for QSO J0831+5245 and QSO B0350-073, and that no redshift uncertainties were reported for QSO B0422+004, we assume that all redshift uncertainties are negligible.

The conformal time for the emission event from a distant quasar at redshift $z$ may be written \cite{Friedman2013}
%%%%%%%
\begin{equation}
\eta_k (z) = \int_0^{1 / (1 + z)} \frac{ da}{a^2 E (a) } ,
\label{etaedef}
\end{equation}
upon using $a_e = 1 / (1 + z)$ from Eq.~(\ref{zdef}) (and recalling our convention that $a(t_0) = 1)$. Using $d\eta = H_0 dt / a(t)$, we may similarly compute the (cosmic time) lookback time to the emission event from today as
%%%%%%%%
\begin{equation}
t^{k}_{\rm lb} (z) = \int_{1 / (1 + z)}^1 \frac{ da}{a H(a)} ,
\label{tlbdef}
\end{equation}
again using $a_e = 1 / (1 + z)$. In \Cref{tab:quasars} we list the quasars used in pairs 1 and 2 for $k = {\cal A,B}$, their measured redshifts, $z$, and the corresponding values of 
$\eta_k (z)$
and 
$t^{k}_{\rm lb} (z)$
for each quasar. The present age of the universe corresponds to 
$\eta_0 = \eta (z=0) = 3.20$,
and the lookback time to the hot big bang is $t_{\rm lb} (\infty) = 13.80$ Gyr. 

Neglecting (for the moment) any possible effects from inhomogeneities along the lines of sight between the quasar emission events and our receipt of the cosmic photons on Earth, we assume that any local-realist mechanism that could have engineered the observed violations of the Bell-CHSH inequality must have acted within the past lightcone of either quasar emission event. Only within those spacetime regions could the local-realist mechanism have altered or previewed the bit that we would later receive on Earth, and shared that information (at or below the speed of light) with other elements of our experimental apparatus, such as the source of entangled particles or the detectors on the other side of our experiment \cite{handsteiner2017a,Gallicchio2014}. For pair 1, any such local-realist mechanism is constrained to have acted no more recently than $t_{\rm lb} = 7.78$ Gyr ago, while for pair 2 the constraint is $t_{\rm lb} = 3.22$ Gyr ago.

We may further characterize the space-time region within which any local-realist mechanism could have acted in order to produce the observed violations of the Bell-CHSH inequality. That region consists of the union of the past lightcones from the quasar emission events utilized for a given experimental run, $V^{(4)}_Q$, which we may compare with the space-time 4-volume of the past lightcone of the experiment itself, $V^{(4)}_{\rm exp}$. To calculate $V^{(4)}_Q$, we must consider the 4-volume of the past light cone from each emission event and subtract the 4-volume of those light cones' intersection.

We calculate the 4-volume contained within the past light cone of a quasar emission event by integrating the invariant volume element $dV = \sqrt{-g} \, d^4 x$ over the region bounded by past-directed null geodesics extending from the quasar emission event, where $g = {\rm det} [g_{\mu\nu} (x)]$ is the determinant of the space-time metric. As we saw above, null geodesics take the form $d \eta = d \chi$ in the coordinates of Eq.~(\ref{dseta}). 
Taking the spatial origin to lie along the worldline of quasar $k$, the 4-volume of the past light cone from emission event ${\cal Q}_k$ may be evaluated as
%%%%%%%
\begin{eqnarray}
V^{(4)} (\eta_k) &=& R_0^4 \int \Theta (\eta_k - \eta - \chi ) \, a^4 (\eta) \chi^2 \sin\theta d \eta d \chi d \theta d\phi \nonumber \\
&=& \frac{4 \pi R_0^4}{3} \int_0^{\eta_k} d\eta \> a^4 (\eta) (\eta_k - \eta)^3 ,
\label{V4k}
\end{eqnarray}
where $\Theta (x)$ is the Heaviside step function. In a 
%spatially flat 
FLRW universe, the most recent (conformal) time at which the past light cones from emission events ${\cal Q}_A$ and ${\cal Q}_B$ overlap is given by \cite{Friedman2013}
%%%%%%
\begin{equation}
\eta_{AB} = \frac{1}{2} \left( \eta_A + \eta_B - \chi_L \right).
\label{etaABdef}
\end{equation}
Here $\chi_L$ is the comoving spatial distance between the worldlines of quasars $A$ and $B$, which, in a spatially flat universe, is given by
%%%%%%%
\begin{equation}
\chi_L = \sqrt{ \chi_A^2 + \chi_B^2 - 2 \chi_A \chi_B \cos \alpha } ,
\label{chiLdef}
\end{equation}
where $\alpha$ is the angle between the quasars as seen from Earth, and \cite{Friedman2013}
%%%%%
\begin{equation}
\chi_k (z) = \int_{1 / (1 + z)}^1 \frac{da}{a^2 E (a) } .
\label{chik}
\end{equation}
Along any $\eta = {\rm constant}$ surface, with $0 \leq \eta \leq \eta_{AB}$, the past light cones from emission events ${\cal Q}_A$ and ${\cal Q}_B$ appear as three-dimensional spheres that partially overlap. In Euclidean space, the (spatial) three-volume of the intersection region of two spheres of radii $r_1$ and $r_2$, with distance between their centers $d$, is given by \cite{Kern38}
%%%%%%%%%%
\begin{equation}
V^{(3)}_I = \frac{ \pi}{12 d} \left( r_1 + r_2 - d \right)^2 \left[ d^2 + 2d (r_1 + r_2) - 3 (r_1 - r_2)^2 \right] .
\label{V3Idef}
\end{equation}
Upon substituting $d \rightarrow \chi_L$, $r_1 \rightarrow \eta_A - \eta$, and $r_2 \rightarrow \eta_B - \eta$, making use of Eq.~(\ref{etaABdef}) for $\eta_{AB}$, and performing some straightforward algebra, we find the space-time 4-volume of the intersection region of the past light cones from emission events ${\cal Q}_A$ and ${\cal Q}_B$ to be
%%%%%%%%%
\begin{eqnarray}
V^{(4)}_I &(\eta_A,& \eta_B, \alpha) \nonumber \\
&=& 4 \pi R_0^4 \int_0^{\eta_{AB}} d\eta \> a^4 (\eta) \left[\frac{ 1}{3} (\eta_{AB} - \eta)^3 \right. \nonumber \\
&\quad& \quad + \left. (\eta_{AB} - \eta)^2 \left( \frac{ \chi_L^2 - (\eta_A - \eta_B )^2}{4\chi_L } \right) \right] .
\label{V4I}
\end{eqnarray}
The union of the past light cones from emission events ${\cal Q}_A$ and ${\cal Q}_B$ therefore has the 4-volume
%%%%%%%
\begin{equation}
V_Q^{(4)} (\eta_A, \eta_B, \alpha) = V^{(4)} (\eta_A) + V^{(4)} (\eta_B) - V_I^{(4)} (\eta_A, \eta_B, \alpha) ,
\label{VQdef}
\end{equation}
while the 4-volume of the past light cone of our experiment is given by $V_{\rm exp}^{(4)} = V^{(4)} (\eta_0)$. 

\begin{figure*}
\centering
\begin{tabular}{@{}c@{}c@{}}
\includegraphics[width=0.47\textwidth]{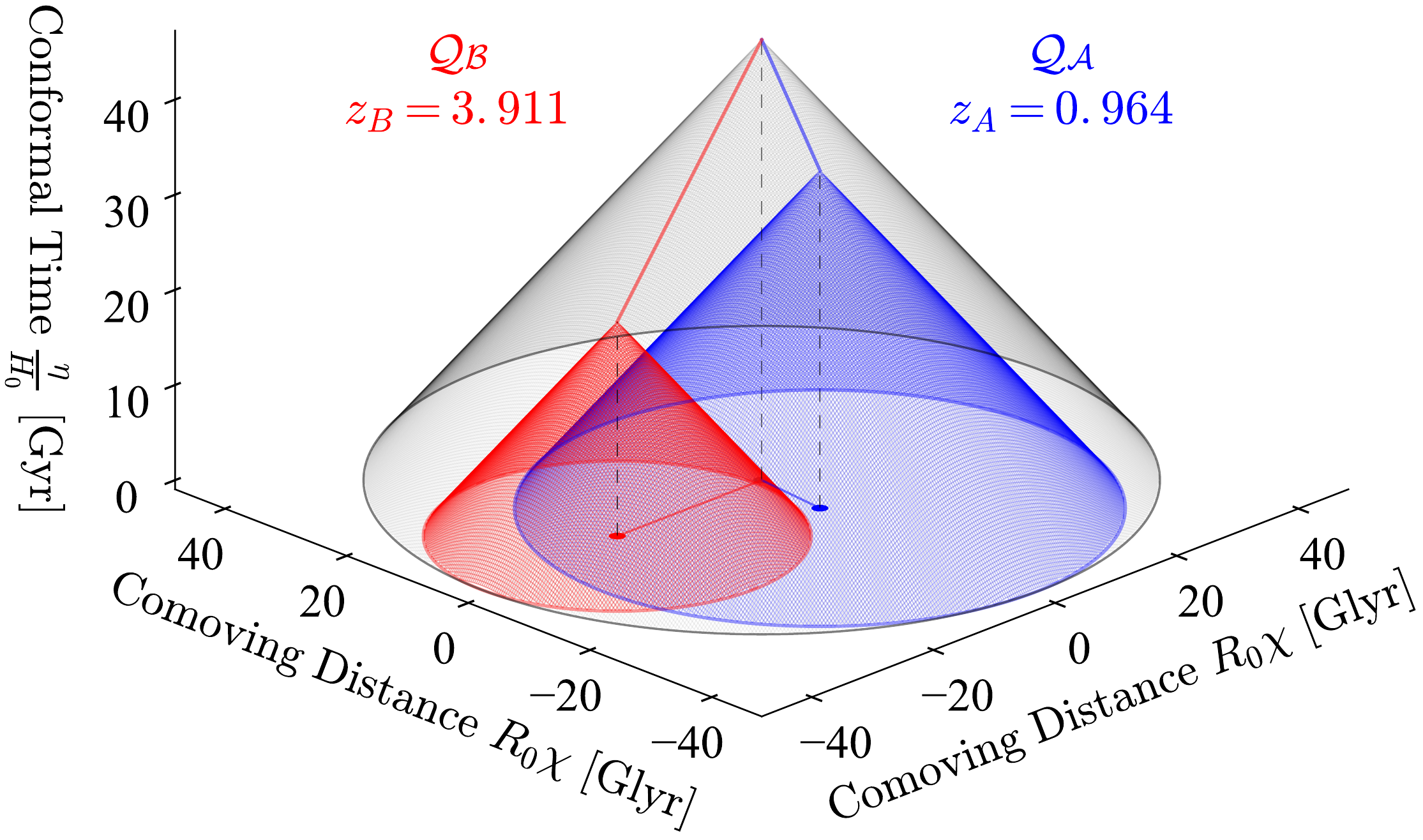} & \quad\quad \includegraphics[width=0.47\textwidth]{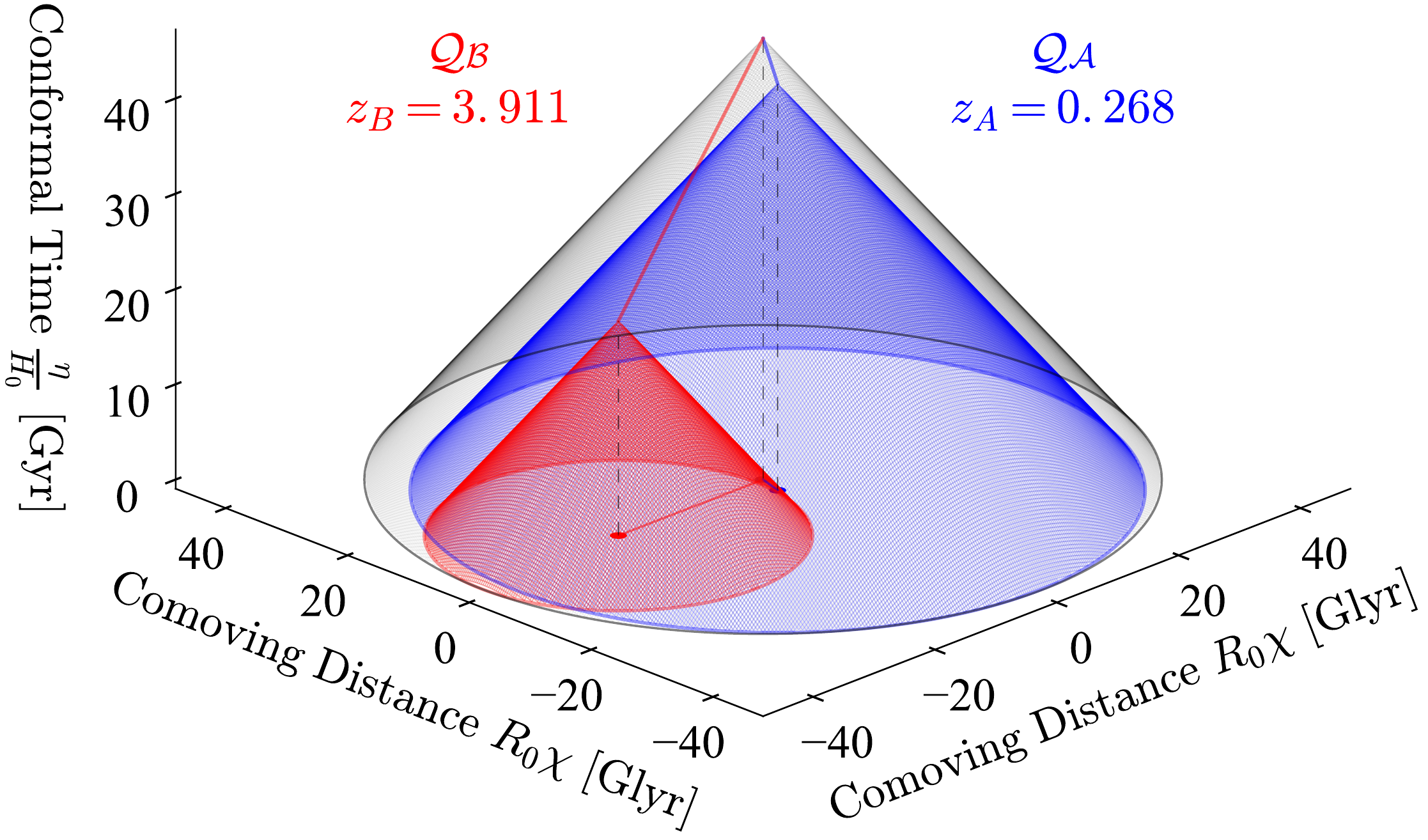} \\
\includegraphics[width=0.47\textwidth]{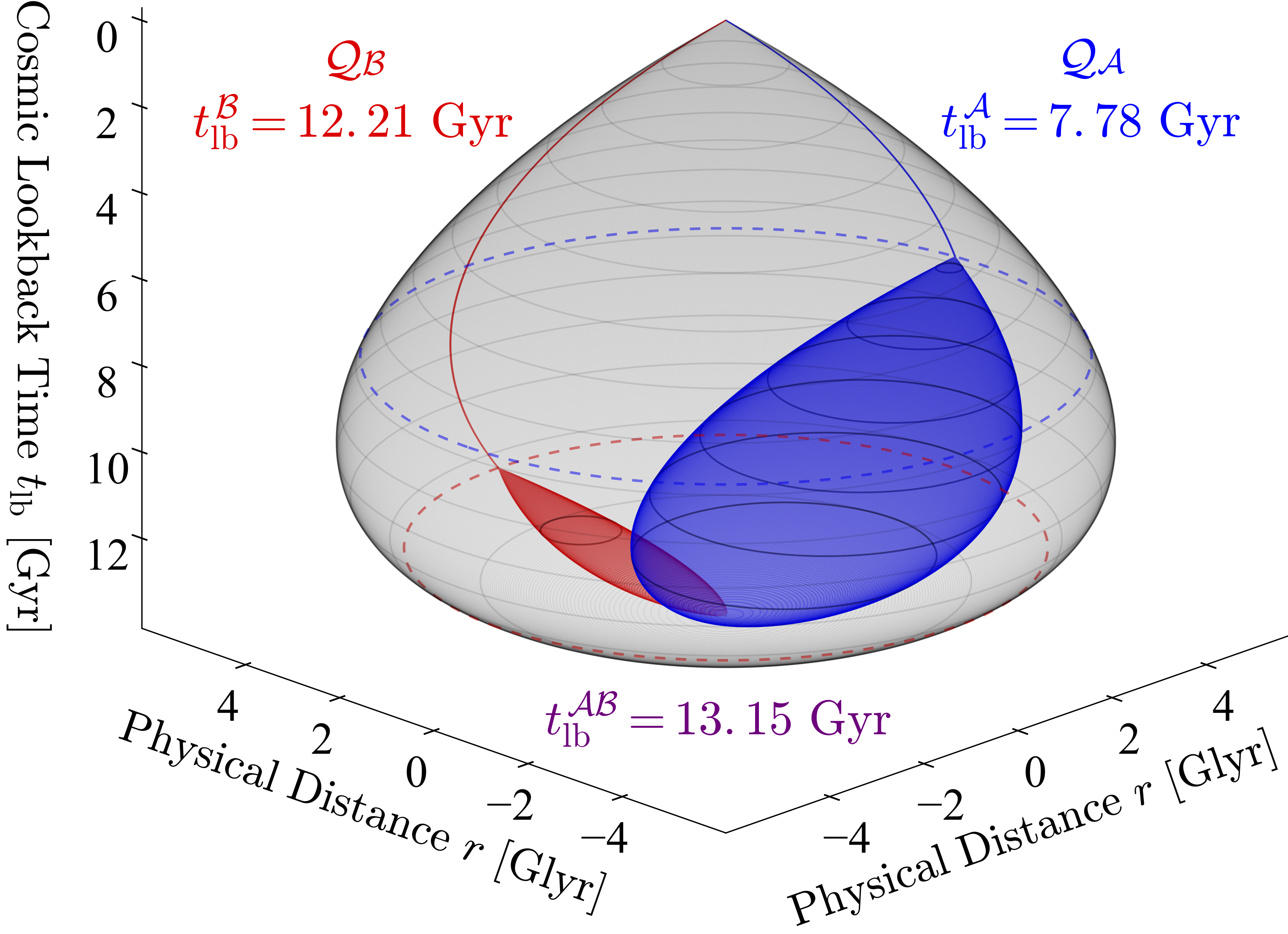} & \quad\quad 
\includegraphics[width=0.47\textwidth]{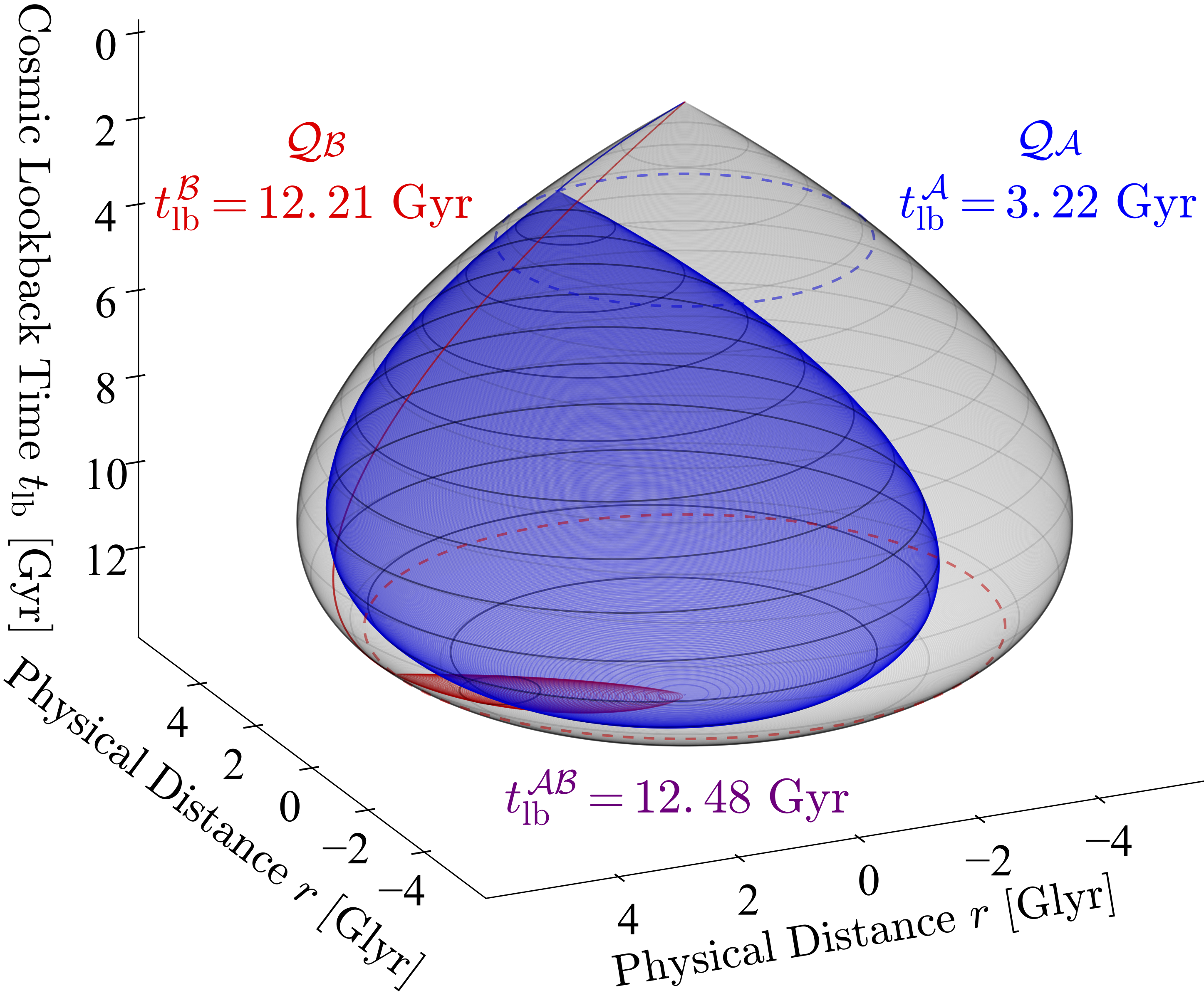} \\
\end{tabular}
\caption{\small 
$(2+1)$D space-time diagrams (with one spatial dimension suppressed) for the quasars in pairs 1 ({\it left}) and 2 ({\it right}), showing the past light cones from quasar emission events 
${\cal Q}_A$ (blue), ${\cal Q}_B$ (red), and the past light cone of our experiment (gray). Top row figures show (rescaled) conformal time $\eta / H_0$ and comoving distance $R_0 \chi$; bottom row figures show cosmic lookback time $t_{\rm lb}$ and physical distance $r = R_0 a(t) \chi$. In conformal coordinates, the big bang singularity is mapped to the surface $\eta = 0$, much as the Earth's poles in a Mercator projection are depicted as lines of comparable length to the Earth's equator. The angle between the red and blue vectors in the $\eta=0$ plane in the conformal diagrams (top row) represents the quasar pair's angular separation on the sky, given by $\alpha = 83.81^{\circ}$ and $\alpha = 72.84^{\circ}$ for pairs 1 and 2, respectively. The bottom-row plots show the lookback times for each emission event ($t_{\rm lb}^k$) and the lookback time to when their past light cones most recently intersected ($t_{\rm lb}^{\cal AB}$). Note that the current age of the universe is $t_0=13.80$ Gyr. The teardrop shape of the past light cones when plotted in $(t, r)$ coordinates arises from the varying expansion rate of the universe over time, given by the scale factor $a(t)$. The paths traveled by the quasar photons along the surface of the gray past light cone of the experiment are shown in red and blue. In each figure, a local-realist mechanism could have exploited information from the red and/or blue regions (and their overlap) to engineer the observed violations of the Bell-CHSH inequality. The gray 
regions outside of the red and blue regions are {\it excluded} from any such local-realist explanation. Assuming negligible uncertainties for the reported redshifts, these excluded regions amount to $F_{\rm excl} = 96.0\%$ (pair 1) and $63.5\%$ (pair 2) of the total space-time 4-volume within the past light cone of our experiment, spanning all of cosmic history since the big bang.}
\label{fig:VolumeExcluded}
\end{figure*}

\begin{figure}
\centering
\includegraphics[width=0.49\textwidth]{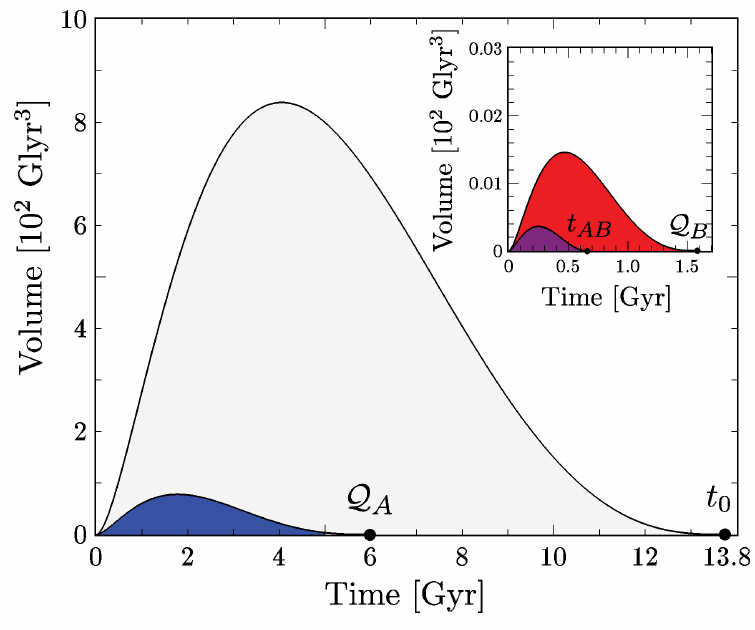}
\caption{\small Spatial volumes as a function of cosmic time $t$ for the relevant past light cones for pair 1. The gray area shows the full past light cone of our experiment, and the blue area shows the past light cone of the emission event ${\cal Q}_A $ at $t_{\cal A} = 6.02$ Gyr ($t_{\rm lb}^{\cal A} = 7.78$ Gyr ago). The spacetime volume of this region is only $4.0\%$ of the full past light cone. The insert shows emission event ${\cal Q}_B$, at $t_{\cal B} = 1.59$ Gyr ($t_{\rm lb}^{\cal B} = 12.21$ Gyr ago). The space-time volume of its past light cone (red) is only $0.020\%$ of the full past light cone. The past light cones of ${\cal Q}_{\cal A}$ and ${\cal Q}_{\cal B}$ last intersected at $t_{\cal AB} = 0.65$ Gyr ($t_{\rm lb}^{\cal AB} = 13.15$ Gyr ago), and the space-time volume of their intersection (purple) is only $0.0023\%$ of the total. 
}
\label{fig:VolumeCompare}
\end{figure}

For an experimental run using a pair of quasars with redshifts $z_A$, $z_B$, and relative angle $\alpha$, the space-time region that is {\it excluded} from playing any role in an explanation based on a local-realist mechanism is given by $V_{\rm exp}^{(4)} - V_Q^{(4)}$. As a fraction of $V_{\rm exp}^{(4)}$, this may be written
%%%%%%%%%
\begin{equation}
F_{\rm excl} = 1 - \left( \frac{ V^{(4)}_Q (\eta_A, \eta_B, \alpha) }{ V^{(4)} (\eta_0) } \right) .
\label{Fdef}
\end{equation}
The relative angle (as seen from Earth) between the quasars in pair 1 is $\alpha = 83.81^\circ$, and for the quasars of pair 2 is $\alpha = 72.84^\circ$. 
Given the redshifts for each quasar listed in \Cref{tab:quasars}, we find $F_{\rm excl} = 0.960$ for pair 1, and $F_{\rm excl} = 0.635$ for pair 2. See \Cref{fig:VolumeExcluded} and \Cref{fig:VolumeCompare}.

We may compare these values for $F_{\rm excl}$ with the corresponding values for our Vienna pilot test involving Milky Way stars \cite{handsteiner2017a}. We again use Eq.~(\ref{zdef}) to relate redshift to cosmic scale factor. Since the motions of Milky Way stars are dominated by local peculiar velocities independent of Hubble expansion, astronomers do not measure cosmic-expansion redshifts for Milky Way stars. Nonetheless, we may compute effective values of $z$ with which to parameterize the various emission times. For the nearby sources used in Ref.~\cite{handsteiner2017a}, we may Taylor expand
%%%%%%
\begin{equation}
    a (t_e) = a (t_0) - H_0 (t_0 - t_e) + {\cal O} ( [(t_0 - t_e)/t_H ]^2 ) ,
    \label{aMilkyWay}
\end{equation}
where, as above, $H_0 = \dot{a} (t_0) / a(t_0)$ is the present value of the Hubble parameter, $t_H = H_0^{-1} = 14.43$ Gyr, and we scale $a (t_0) = 1$. Neglecting uncertainties on the measured distances to the Milky Way stars that we used in our pilot test, we have $(t_0 - t_A) = 604$ years and $(t_0 - t_B) = 1930$ years for run 1, and $(t_0 - t_A) = 577$ years and $(t_0 - t_B) = 3624$ years for run 2. Using Eqs.~(\ref{zdef}) and (\ref{aMilkyWay}), these correspond to effective redshifts $z_A = 4.19 \times 10^{-8}$ and $z_B = 1.32 \times 10^{-7}$ for run 1, and $z_A = 4.00 \times 10^{-8}$ and $z_B = 2.51 \times 10^{-7}$ for run 2. Given $\alpha_1 = 119^\circ$ for run 1 and $\alpha_2 = 112^\circ$ for run 2, the times when the past light cones from emission events $A$ and $B$ most recently overlapped were $t_{AB} = 2409$ years ago (run 1) and $t_{AB} = 4039$ years ago (run 2), corresponding to $z_{AB} = 1.67 \times 10^{-7}$ (run 1) and $z_{AB} = 2.80 \times 10^{-7}$ (run 2). Repeating the calculation as above, we then find $F_{\rm excl} = 1.38 \times 10^{-7}$ for run 1, and $F_{\rm excl} = 1.45 \times 10^{-7}$ for run 2. In other words, the Vienna pilot test excluded about one hundred-thousandth of one percent of the relevant space-time volume, compared to the exclusion of $96.0\%$ (pair 1) and $63.5\%$ (pair 2) achieved in the present experiment.

\section{Effects of Inhomogeneities along the Line of Sight} 

Our discussion to this point has assumed that the quasar photons were emitted from point-like astronomical objects. In reality, quasars are large, messy objects; a given photon may be subject to complicated interactions involving optically thick atmospheres before escaping from the quasar. To address such scenarios, we consider the ``effective emission time" to be the latest time 
%$\eta_{q_k}$
$\eta_{k}$
that any local interactions associated with the quasar could have altered the wavelength of the photon. Any corrections arising from strong electromagnetic fields, plasma effects, or related 
atmospheric phenomena near the quasar would affect a precise calculation of 
%$\eta_{q_k}$
$\eta_{k}$
by some small quantity (with cosmic-time values for the corrections 
%$\Delta t \ll t_{\rm lb}$).
$\Delta t \ll t^{k}_{\rm lb}$). 
Compared to our simple estimates of 
%$\eta_{q_k} (z)$
$\eta_{k}(z)$
and 
%$t_{\rm lb} (z)$
$t^{k}_{\rm lb} (z)$
based on the measured redshifts of the quasars, any such corrections would be indiscernible, given 
$t^{k}_{\rm lb} (z) \sim {\cal O} (1 - 10)$ Gyr for the quasars used in pairs 1 and 2.

Atmospheric or related interactions at the quasar could introduce delays between the arrival at Earth of the causal wavefront from the emission event of a given photon and the receipt of that photon on Earth. In principle, a local-realist mechanism could therefore exploit information about the wavelength of the incoming quasar photon prior to its detection, in order to engineer the observed violations of the Bell-CHSH inequality. However, any such advanced signal about the incoming quasar photon would need to be correlated with the detector-setting photon itself, and therefore the information carried by the advanced signal must also have originated within the past light cone of the quasar emission event ${\cal Q}_k$, with effective emission time 
$\eta_{k}(z)$.

Similar considerations apply to the case in which photons from a given quasar are subject to strong gravitational lensing en route from the quasar to Earth. For example, it is known that light from quasar QSO J0831+5245 (which we used to determine Bob's settings in pairs 1 and 2) is lensed by a large, intervening mass \cite{Egami2000,Oya13}, producing multiple images of the original quasar as seen from Earth. The multiple images arise from different paths that quasar photons take between the lens and Earth. In the case of this particular quasar, it has been estimated that the distinct paths correspond to a difference in arrival times at Earth of as much as 1 day \cite{Oya13}.

\begin{figure}
\centering
\includegraphics[width=0.49\textwidth]{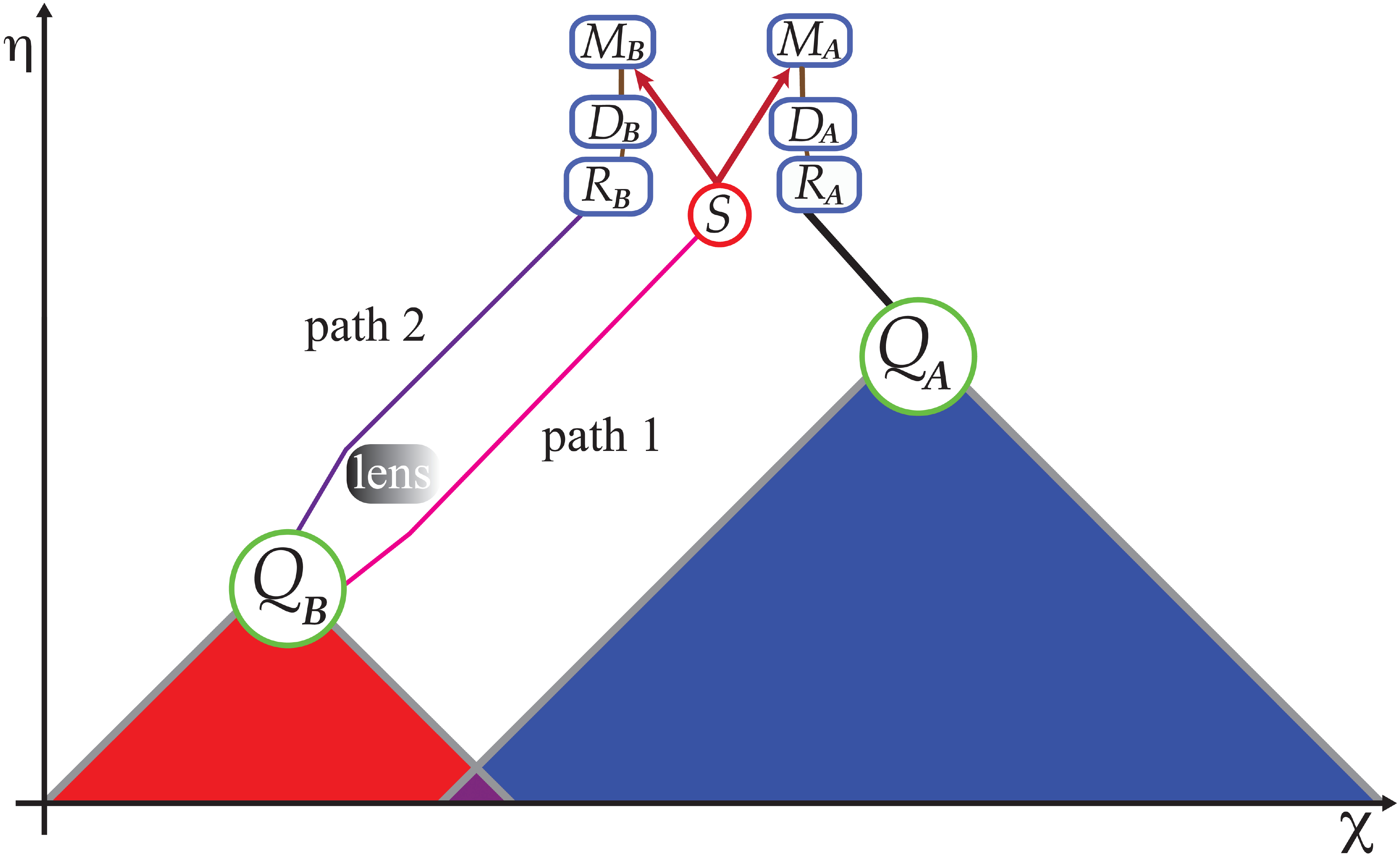}
\caption{\small 
A scenario in which light from quasar emission event ${\cal Q}_B$ is lensed by a large, intervening mass between the quasar and Earth. The lens can produce multiple images of the quasar as seen from Earth. In the scenario shown, photons following path 1 arrive at Earth before photons that follow path 2. However, to be valuable to a local-realist mechanism, the information that arrives along path 1 must be correlated with information that originates within the past light cone of the effective emission event ${\cal Q}_B$ (red or purple regions).
}
\label{fig:lensing}
\end{figure}

Given the delay in arrival times, it is possible (in principle) that a local-realist mechanism could receive information from a short-path photon about the wavelength of a long-path photon before the latter arrives at Earth, and exploit that advanced signal to engineer the observed violations of the Bell-CHSH inequality. Much like the case of atmospheric delays at the quasar itself, however, any information of value to the local-realist mechanism would need to have originated within the past light cone of the emission event, with effective emission time, 
$\eta_{k}(z)$.
Such scenarios are therefore constrained to the same space-time regions described above. See \Cref{fig:lensing}.

One scenario in which gravitational lensing {\it could} affect our conclusions would arise if the wavelength of a quasar photon could somehow be measured without altering the photon's wavelength or trajectory. (No such measurement is possible according to quantum mechanics, but a local-realist mechanism, by design, is meant to be distinct from quantum mechanics.) If an ``in-flight" measurement were possible, then a local-realist mechanism could potentially measure the wavelength of a quasar photon as it approaches the gravitational lens, and arrange for information about that photon's wavelength to arrive at Earth via a short path from the lens, rather than a long path.
In such a scenario, the most recent time by which the local-realist mechanism would need to have acted would be bounded by the time the quasar photons reach the lens, which is more recent than the emission-time from the quasar. (We do not consider a scenario in which the local-realist mechanism could {\it alter} the wavelength of the quasar photon without changing its trajectory; any such mechanism would violate the conservation of energy and momentum.)

In the case of the lensed quasar in our study, the intervening lens has an estimated redshift $z_{\rm lens} \simeq 3$ \cite{Egami2000}. This corresponds to a lookback time from Earth of $t_{\rm lb} (z_{\rm lens} ) \simeq 11.6$ Gyr ago, compared to the quasar emission time $t_{\rm lb}^{\cal B} = 12.21$ Gyr ago. The time at which photons arrive at the lens is considerably earlier than the emission times from either of the quasars with which this quasar was paired ($t_{\rm lb}^{\cal A} = 7.78$ Gyr for pair 1 and $t_{\rm lb}^{\cal A} = 3.22$ Gyr for pair 2), so even such an in-flight measurement scenario would have no impact on our overall conclusions. Likewise, because the fraction of the relevant space-time 4-volume is dominated by the volume of the past light cone of the more recent emission event, re-calculating $F_{\rm excl}$ using $z_{\rm lens} \simeq 3$ rather than $z_B = 3.911$ yields virtually no change compared to the values computed above: $F_{\rm excl, lens} = 0.960$ for pair 1, and $F_{\rm excl, lens} = 0.635$ for pair 2. 

Photons from distant quasars can be affected in other ways between emission and detection, beyond gravitational lensing. In particular, the intergalactic medium can affect the quasar spectra observed on Earth. Quasar sources typically have strong emission at the Lyman-$\alpha$ line, which, in the laboratory frame, corresponds to $\lambda_{\rm emit}^\alpha = 121.6$ nm (deep in the ultraviolet). However, en route, photons from high-redshift quasars encounter clouds of neutral hydrogen gas, which preferentially absorb photons at the Lyman-$\alpha$ wavelength --- for quasar photons that have been redshifted during their travel to $121.6$ nm by the time they encounter the gas cloud. Photons from very distant quasars can encounter multiple gas clouds en route to Earth, resulting in a dense ``Lyman-$\alpha$ forest" of absorption lines at wavelengths shorter than $\lambda_{\rm obs}^\alpha = (1 + z) \lambda_{\rm emit}^\alpha$, where $z$ is the redshift of the original quasar emission \cite{RauchQSOreview,McDonaldLyman}.

In our experiment, Alice's receiving station observed quasars at redshifts $z_A = 0.964$ (pair 1) and $z_A = 0.268$ (pair 2). Hence the Lyman-$\alpha$ forest affected quasar photons that would have been received with wavelengths $\lambda_{\rm obs} < 238.8$ nm (pair 1) and $\lambda_{\rm obs} < 154.2$ nm (pair 2). But, as discussed further below, the detectors of our cosmic random number generators (CRNGs) were largely insensitive to $\lambda_{\rm obs} < 400$ nm. Hence selective absorption by the Lyman-$\alpha$ forest would have had no observable effect for either of the ``quasar $A$" sources in our experiment. This also means that effects from transmission through the intergalactic medium could not have introduced correlations between the detected photons from quasars $A$ and $B$, because any effects on the quasar-$A$ photons would have fallen outside the sensitivity range of our detectors.

On the other hand, at Bob's receiving station we used a more distant quasar, with $z_B = 3.911$. The photons from this distant quasar certainly could have been affected by selective absorption within the Lyman-$\alpha$ forest, for observed wavelengths $\lambda < \lambda_{\rm obs}^\alpha = (1 + z_B) \lambda_{\rm emit}^\alpha = 597.2$ nm. Our 
CRNGs used dichroic filters to distinguish ``red" from ``blue" astronomical photons on either side of $\lambda_{\rm filter} = 630$ nm. Therefore the entire range of photons from quasar $B$ that could have been affected by the Lyman-$\alpha$ forest falls within the ``blue" channel. Akin to the gravitational-lensing scenario described above, one may imagine some local-realist ``conspiracy" that used the Lyman-$\alpha$ forest to purposefully alter the spectra that would be received at Bob's station (by suppressing ``blue" photons), and/or that could have ``alerted" other elements of our experimental apparatus with a preview of the patterns in the upcoming sequence of ``red" and ``blue" detections at Bob's station.

However, given our detectors' sensitivity for $\lambda \geq 400$ nm, the {\it closest} gas cloud that could have affected the observable portion of the spectrum from quasar $B$ would be at some redshift $z_{\rm cloud}$ such that $400 \, {\rm nm} = (1 + z_{\rm cloud}) \lambda_{\rm emit}^\alpha$, or $z_{\rm cloud} = 2.29$, corresponding to a lookback time at which the quasar photons encountered that cloud of $t_{\rm lb}^{\rm cloud} = 10.93$ Gyr ago. This falls considerably longer ago than the lookback times to the emission events from quasar $A$: $t_{\rm lb}^{\cal A} = 7.78$ Gyr ago for pair 1, $t_{\rm lb}^{\cal A} = 3.22$ Gyr ago for pair 2. Therefore, any ``conspiracy" that might have occurred as recently as $t_{\rm lb}^{\rm cloud} = 10.93$ Gyr ago is consistent with our overall conclusion: we have constrained any such local-realist mechanisms to have occurred no more recently than $t_{\rm lb}^{\cal A}$. Likewise, we may calculate the excluded space-time volume fraction under the assumption that some ``conspiracy" occurred at a gas cloud at $z_{\rm cloud} = 2.29$, by substituting $z_B \rightarrow z_{\rm cloud}$. Repeating the calculation as above, we then find $F_{\rm excl} = 0.958$ rather than $0.960$ for pair 1, whereas for pair 2, we find $F_{\rm excl} = 0.635$, unchanged from our original calculation.

\section{Quasar Selection}

\begin{figure}
\centering
\includegraphics[width=0.49\textwidth]{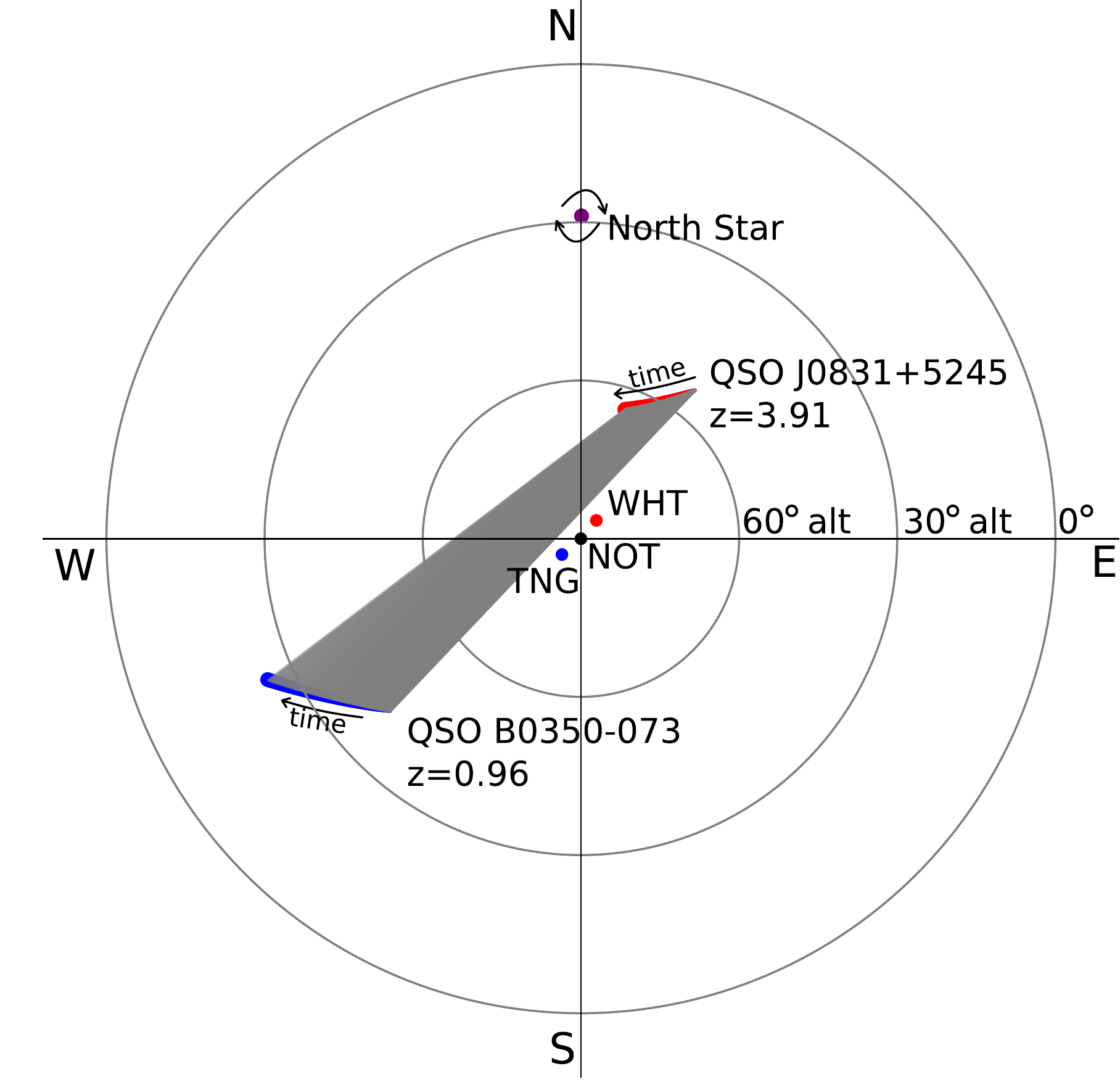}
\caption{\small 
Overhead view of the orientation of quasars in the sky with respect to the source at NOT and the receivers at TNG and WHT on 2018-01-11 between 00:00 and 01:30 UTC, ending with the pair 1 measurement.}
\label{fig:pair1_overhead}
\end{figure}

In anticipation of our limited observing opportunities at the observatory, we searched for quasar pairs whose space-time arrangement would exclude the largest fraction of the 4-volume of the past light cone of our experiment, while maintaining a sufficiently high signal-to-noise ratio to yield a statistically significant result, given constraints on telescope time.
(Following Refs.~\cite{handsteiner2017a,Kofler16,leung2017}, we discuss requirements for signal-to-noise below.)
We started with a database of the $\sim$62\,000 quasars from the Simbad database that had a Sloan r$^{\prime}$-band magnitude brighter than 19. We cross-referenced this to the SDSS DR14 database, taking the most conservative of their redshift estimates (or other redshifts from Simbad and the literature, where SDSS redshifts were not available). We considered
only quasars that were the brightest for their distance within each $5^\circ$ by $5^\circ$ patch of sky. This left $\sim$4\,000 quasars. 
For each pair of these, and for every minute of allotted telescope time, we simulated the number of quasar and skyglow photons that each telescope's detectors would record when pointed at the relevant instantaneous elevation angle $>25^{\circ}$ and observed through the associated airmass. Figure \ref{fig:pair1_overhead} shows the path on the sky that the quasars of pair 1 took during a period of 1.5 hours on one of our observing nights;
our experiment with pair 1 was conducted near the end of that window. 

To estimate the rate at which entangled photon coincidences would accumulate, 
we required that the relevant signal-to-noise threshold be exceeded in each of the red/blue detector-setting ports for both quasars, and that detectors be triggered while both quasars were in causal alignment with respect to the experimental stations such that $\tau^k_{\rm valid} (t)>0$ 
for both sides $k=A, B$. 
This ensured that we only included entangled-photon coincidences while closing the locality loophole and ensuring the signal-to-noise was sufficiently high.

For a given experimental visibility, entangled-photon coincidence rate, and quasar-photon rate, we estimated the statistical significance we could achieve during the time window while all these conditions were met. 
Then we chose the highest-redshift
pairs with the largest observable angular separations that 
our simulations predicted could
yield significant results in the time allotted.

For the best of these candidates, we further required each object to have published spectra and verified redshifts (either SDSS DR14 or elsewhere in the literature). We manually vetted these to ensure they were legitimate quasar spectra and accurately estimated redshifts, and not, for example, misidentified stellar spectra of $r^{\prime}$$\sim$$13$-$19$ magnitude stars (which the SDSS algorithms sometimes misclassified as extragalactic objects with redshifts $z>0.1$ \cite{Paris17}). These cases were non-negligible as contaminants to the subset of the SDSS DR14 database that our software preferentially selected, so the manual checks were required before choosing final target quasars to observe. For each of the four initially scheduled time slots on the telescopes, we performed this procedure to select $\sim$10-20 vetted targets, yielding several possible pairs that would be optimal at the beginning of the scheduled observation window. The best pairs balanced the trade-off between the largest excluded cosmological 4-volume and the highest statistical significance that could be expected to accumulate within the relevant time window. 

Although we were originally scheduled for observation windows over the course of four consecutive nights, due to bad weather and technical problems, we were only able to conduct experiments during the last of our scheduled sessions (early in the morning of 11 January 2018).

\section{Experimental Details}
We used Type-0 spontaneous parametric down conversion in a \SI{30}{\milli\metre}  periodically poled $\rm KTiOPO_4$ (ppKTP) crystal placed in a Sagnac-interferometer loop to produce entangled photon pairs. The crystal was bi-directionally pumped by a grating-stabilized \SI{405}{\nano\metre} laser to produce pairs of horizontally polarized down-converted photons at wavelengths of \SI{773.6}{\nano\metre} and \SI{850}{\nano\metre} with a spectral bandwidth of \SI{\approx 2.5}{\nano\metre} FWHM. The entangled state was rotated near the maximally entangled Bell state $|\psi^{-}\rangle = \frac{1}{\sqrt2}(|HV\rangle - |VH\rangle)$ with 
fiber polarization controller paddles. The relative phase was adjusted by the polarization of the pump beam by optimizing the Bell test visibility. 
We locally measured heralding efficiencies of 31\% and 41\%, which differed by about 1\% before and after the experiment, confirming that these parameters remained stable over the duration of the measurement.
For pair 1 (pair 2) the duty cycle of Alice's and Bob's measurements - i.e. the temporal sum of used valid setting intervals divided by the total measurement time per run - were \SI{0.32}{\percent} (\SI{1.17}{\percent}) and \SI{1.62}{\percent} (\SI{0.96}{\percent}), respectively, resulting in a duty-cycle for valid coincidence detections between Alice and Bob of \num{5.2d-5} (\num{1.1d-4}).

\section{Dichroic Selection of CRNGs}
As emphasized in our previous work \cite{handsteiner2017a,leung2017}, it is critical to select appropriate dichroic filters for generating astronomical random bits on the basis of their wavelength. The filters should be  chosen to minimize the predictability of the random bits generated. This means minimizing cross-talk between the ``red'' and ``blue'' detection channels, and choosing the infrared cutoff of our red band to be opaque to skyglow, which at the Roque de los Muchachos observatory increases rapidly with increasing wavelength over the range $\sim 700-1000$ nm \cite{benn1999la} due to transitions in the rovibrational states of OH radicals which are abundant in the upper atmosphere \cite{meinel1950}.

We employed a system of four dichroic elements which define our detection bands as shown in \Cref{fig:twopanel}. First, a longpass dichroic beamsplitter (BS) is used to reflect incoming light with $\lambda \lesssim 635$ nm and send it towards our ``blue'' detector. After the dichroic beamsplitter, we place additional shortpass (SP1: $\lambda < 620$ nm) and longpass filters (LP1: $\lambda > 637$ nm) in the blue and red output ports to reject photons near the transition wavelength, which may go either way at the dichroic beamsplitter. This way, the fraction of red photons detected in the blue arm and vice versa  is negligible ($f_w < 2\times10^{-5}$). This represents a significant improvement over our previous experiment, in which the wrong-way fractions for each detector were $f_w \sim {\cal O} (10^{-2})$~\cite{handsteiner2017a}. Finally, after the longpass filter in the transmitted (red) output port, we define the long-wavelength cutoff of our red band with a shortpass filter (SP2) at $\approx 745$ nm, a wavelength chosen to optimize the trade-off between rejection of infrared skyglow and transmission of quasar photons.

In order to make the selection of the dichroic elements BS, SP1, LP1, and SP2, we began with a hand-prepared list of available filtersets (BS, SP1, LP1) whose wavelengths were compatible as well as a list of available filters SP2. Then, for every combination of (BS, SP1, LP1) and SP2, we computed the signal-to-noise ratio in the red and blue channels up to an unknown overall constant which varies from quasar to quasar. We chose the final filter combination that yielded a strong signal-to-noise ratio for all three quasars observed. For every observation except that of QSO B0422+004, our first-principles computation of the red-blue imbalance agreed with the measured count rates to within 11\%. For the violently-variable BL Lac object QSO B0422+004, our observation was redder than predicted by our model by a factor of $\approx 1.6$. This is consistent with recent photometric observations, which report a brightening of QSO B0422+004 in the V band by a factor of 2.5 between December 9, 2014, and December 29, 2015~\cite{2015atel6866}. Similar dramatic variations in brightness were also observed in the $1.2-2.2$ micron J, H, and K bands between February 2013 and January 2015~\cite{2014atel5712,2015atel6865,2015atel6971}.

\begin{figure}
    \centering
    \includegraphics[width = 0.5\textwidth]{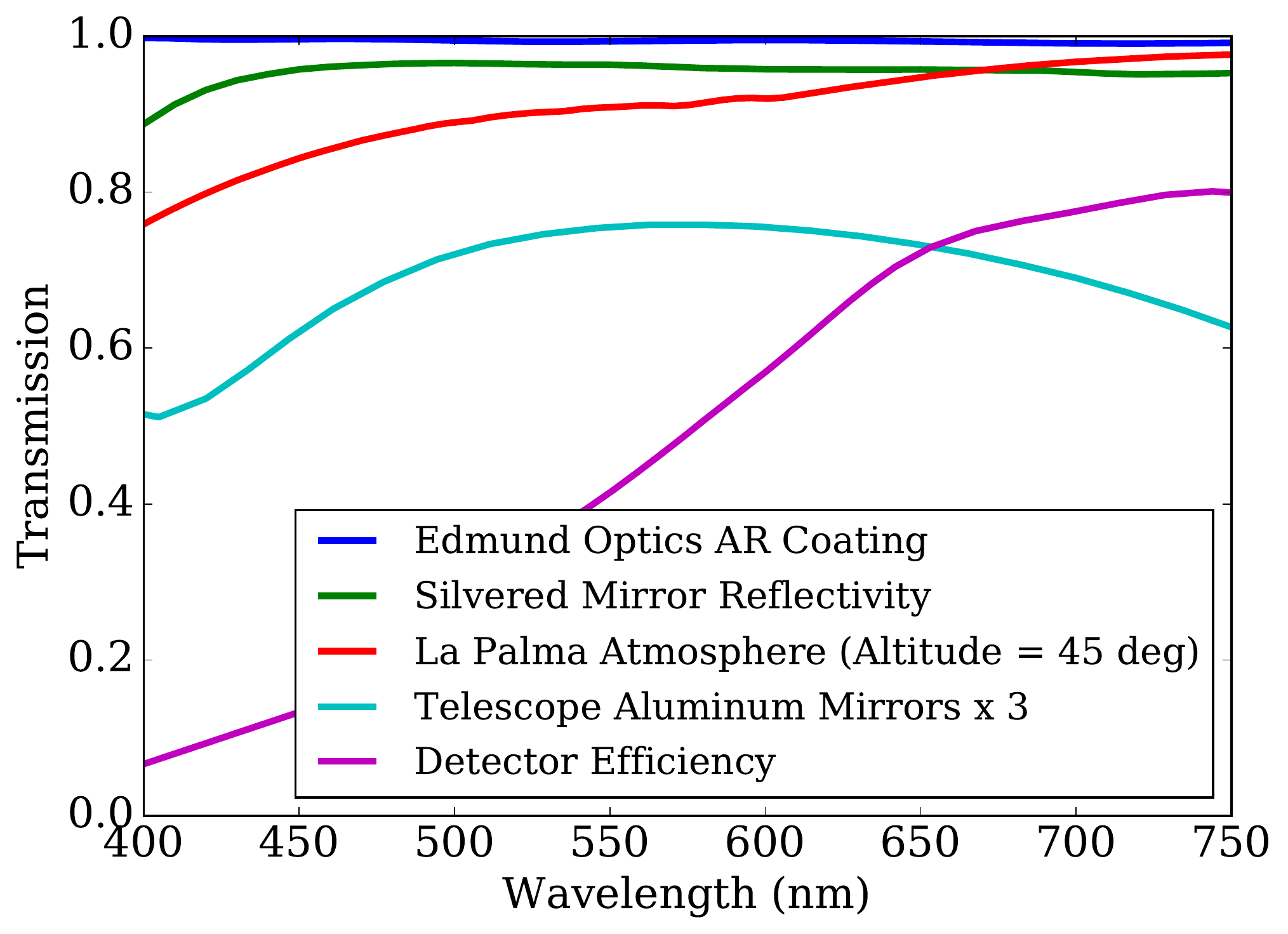}
    \caption{Transmission spectra of optical elements in our Cosmic RNGs. Both the WHT and TNG have three aluminum mirrors, whose cumulative transmission is plotted here.}
    \label{fig:transcomponents}
\end{figure}

\begin{figure}
    \centering
    \includegraphics[width = 0.5 \textwidth]{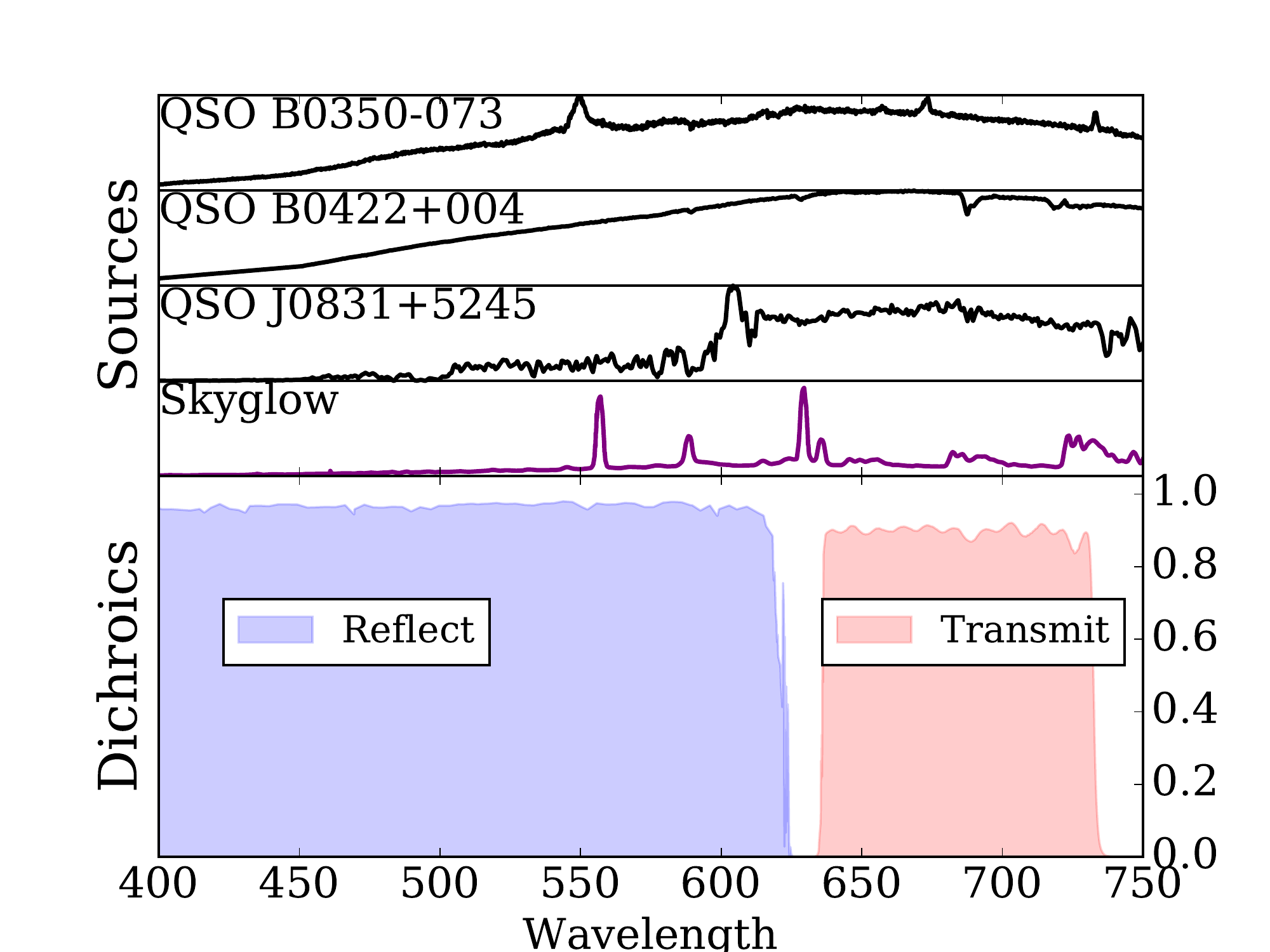}
    \caption{Top: The number distribution of photons from each source, weighted by the transmission of optical elements and the detector efficiency. The quasar spectra are reddened by Rayleigh scattering. Bottom: Our system of four dichroic elements splits incoming photons cleanly into a reflect (blue) channel and a transmit (red) channel. The red channel is shortpassed to reject a large fraction of Meinel rovibrational emissions which grow rapidly at wavelengths longer than $\sim 700$ nm. Quasar spectra were taken from: QSO J0831+5245 \cite{Saturni16}, QSO B0350-073 (SDSS DR14, \cite{Paris17}), and QSO B0422+004 \cite{Sbarufatti06}. The skyglow spectrum is from Ref.~\cite{benn1999la}.
    }
    \label{fig:twopanel}
\end{figure}

The optical efficiency of our CRNGs is shown in \Cref{tab:dc}. It varies from quasar to quasar due to their diverse spectral shape and different observing altitudes.
\begin{table}[]
    \centering
    \begin{tabular}{l  c c }
    \hline
    \hline 
    Path & $\%$ Transmit (red) & $\%$ Reflect(blue) \\
    \hline
    Through atmosphere  & 95-96 & 86-89 \\
    Through all optics  & 38-39 & 20-30 \\
    Detector quantum efficiency & 75-76 &  30-44 \\
    \hline
    \hline
    \end{tabular}
    \caption{The probability that a quasar photon in our transmit/reflect wavelength band is transmitted through the atmosphere, through the telescope/CRNG collection optics, and the probability of registering as an electronic pulse at the APD.}
    \label{tab:dc}
\end{table}
For all computations, we employ a spectral model, formulated in Ref.~\cite{handsteiner2017a}, which takes as input a quasar spectrum (counts per second per wavelength), its observation altitude, and choices for the dichroic elements BS, SP1, LP1, and LP2. Quasar spectra are corrected for atmospheric extinction with tabulated values for atmospheric reddening at the Roque de los Muchachos~\cite{rgo1985atmospheric}.
Both quasar and skyglow spectra are weighted by the transmission curves of the telescopes' aluminum mirror coatings, our lenses' anti-reflection coatings, and our ID120 detectors' quantum efficiency curve. These constituent curves are plotted in \Cref{fig:transcomponents}, and the weighted spectra are plotted in \Cref{fig:twopanel}.

\section{Data Analysis}
\label{sec:data}

In this section we analyze the data from the two experimental runs, which were each conducted early in the morning of 11 January 2018. We make the assumptions of fair sampling and fair coincidences \cite{Larsson2014}. Thus, for testing local realism, all data can be postselected to coincidence events between Alice's and Bob's measurement stations. These coincidences were identified using a time window of $2.66$ ns.

As in Ref.~\cite{handsteiner2017a}, we denote by $N_{ij}^{AB}$ the number of coincidences in which Alice had outcome $A \in \{ + , - \}$ under setting $a_i$ (with $i = 1, 2$) and Bob had outcome $B \in \{ + , - \}$ under setting $b_j$ (with $j = 1,2$). Then the number of all coincidences for settings combination $a_i b_j$ is given by
%%%%%%%%%%%%%
\begin{equation}
N_{ij} \equiv \sum_{A,B = + , -} N_{ij}^{AB} ,
\label{Nijdef}
\end{equation}
and the total number of all recorded coincidences is 
%%%%%%%%%
\begin{equation}
N \equiv \sum_{i,j = 1,2} N_{ij} .
\label{Ntot}
\end{equation}
A point estimate of the joint setting probabilities is given by
%%%%%%%%
\begin{equation}
q_{ij} \equiv p (a_i b_j) = \frac{ N_{ij} }{N} .
\label{qijdef}
\end{equation}
We may then test whether the probabilities $q_{ij}$ can be factorized, that is, whether they can be written (approximately) as
%%%%%%%%%
\begin{equation}
p_{ij} \equiv p (a_i) p (b_j) ,
\label{pijdef}
\end{equation}
where
%%%%%%%%%%%
\begin{eqnarray}
p (a_i) \equiv \frac{ N_{i1} + N_{i2} }{N} &=& q_{i1} + q_{i2} , \nonumber \\
p (b_j) \equiv \frac{ N_{1j} + N_{2j} }{N} &=& q_{1j} + q_{2j} .
\label{papbdef}
\end{eqnarray}
We may also estimate the conditional probabilities for correlated outcomes in which both Alice and Bob observe the same result:
%%%%%%%%%%%
\begin{equation}
p (A = B \vert a_i b_j) = \frac{ N_{ij}^{++} + N_{ij}^{- -} }{N_{ij} } .
\label{pcorrdef}
\end{equation}
If we define the quantity
%%%%%%%%%
\begin{eqnarray}
C &\equiv& - p (A = B \vert a_1 b_1 ) - p (A = B \vert a_1 b_2) \nonumber \\
&\quad& \quad - p (A= B \vert a_2 b_1) + p (A = B \vert a_2 b_2) ,
\label{Cdef}
\end{eqnarray}
then the Bell-CHSH inequality for local-realist models \cite{Clauser1969} takes the form $C \leq 0$ (if one neglects the ``freedom-of-choice" loophole \cite{Kofler16}). We may also construct the correlation functions
%%%%%%%%%%
\begin{equation}
E_{ij} \equiv 2 p (A = B \vert a_i b_j ) - 1 ,
\label{Eijdef}
\end{equation}
in terms of which we may construct the quantity
%%%%%%%%%%%
\begin{equation}
S \equiv \vert E_{11} + E_{12} + E_{21} - E_{22} \vert .
\label{Sdef}
\end{equation}
The Bell-CHSH inequality may then be written $S \leq 2$. The quantities $C$ and $S$ are related as $S = 2 \vert C + 1 \vert$.

The measured coincidence counts $N_{ij}^{AB}$ for pair 1 are 
\begin{equation}
{\bf Pair \>\> 1: \quad\quad}
\begin{array}
[c]{lrrrr}%
ij\;\backslash\;AB & ++ & +- & -+ & --\vspace{0.1cm}\\
11 & 145 & 956 & 1\,229 & 291 \\
12 & 487 & 1\,618 & 2\,417 & 370 \\
21 & 440 & 1\,514 & 1\,399 & 206\\
22 & 3\,229 & 418 & 593 & 2\,321
\end{array} 
\label{run2coinc}
\end{equation}
For pair 1, we have $N = 17,633$ total coincidence counts. For pair 2, we find
\begin{equation}
{\bf Pair \>\> 2: \quad\quad}
\begin{array}
[c]{lrrrr}%
ij\;\backslash\;AB & ++ & +- & -+ & --\vspace{0.1cm}\\
11 & 71 & 1\,007 & 1\,102 & 168 \\
12 & 654 & 1\,394 & \;1\,975 & 494 \\
21 & 315 & 771 & 702 & 186\\
22 & \;1\,975 & 108 & 165 & \;1\,333
\end{array} 
\label{run1coinc}
\end{equation}
For pair 2, we have $N = 12,420$ total coincidence counts. 

Using the measured coincidence counts in Eqs.~(\ref{run1coinc}) and (\ref{run2coinc}), we find the values for the Bell-CHSH parameters $C$ and $S$ of Eqs.~(\ref{Cdef}) and (\ref{Sdef}) as shown in \Cref{tab:CS}. There we also show the visibility fraction, defined as
%%%%%%%%%
\begin{equation}
{\cal V} \equiv \frac{ S}{S_{\rm max}^{QM}} ,
\label{Vdef}
\end{equation}
where $S_{\rm max}^{QM} = 2\sqrt{2}$, the Tsirelson bound \cite{Cirelson80}, is the maximum value that the quantity $S$ may attain according to quantum mechanics.

\begin{table}[!htbp]
    \centering
    \begin{tabular}{ | c | c | c | c | }
    \hline
    Pair & $C$ & $S$ & $ {\cal V}$ \\
    \hline
    1 & $0.3229$ & $2.6457$ & $0.935$ \\
    \hline
    2 & $0.3140$ & $2.6281$ & $0.929$ \\
    \hline
    \end{tabular}
    \caption{\small Values for the Bell-CHSH parameters and the visibility fraction for each experimental run.}
    \label{tab:CS}
\end{table}

\section{Statistical Independence of Settings Choices}
\label{sec:independence}

We first consider whether the joint settings frequencies $q_{ij}$ for the data from pairs 1 and 2 may be factorized. For the individual settings probabilities $p (a_i)$ and $p (b_j)$ defined in Eq.~(\ref{papbdef}), we find the values shown in \Cref{tab:papb}. The joint frequencies $q_{ij}$ and the inferred joint probabilities $p_{ij}$ are shown in \Cref{tab:qijpij}. Note that for each dataset, we find $\sum_{ij} q_{ij} = \sum_{ij} p_{ij} = 1.00$, as required. 

\begin{table}[]
    \centering
    \begin{tabular}{c c l c c c c }
    \hline \hline
    Pair & Side         & ID          & $n_r$ & $s_r$ &  $n_b$ & $s_b$\\
    \hline 
    1    & $\mathcal{A} $ & QSO B0350-073  & 288  &  2094  & 350   &  2774\\
    1    & $\mathcal{B} $ & QSO J0831+5245 & 358  &  9320  & 408   &  5064\\
    2    & $\mathcal{A} $ & QSO B0422+004  & 640  &  3970  & 684   &  3224\\
    2    & $\mathcal{B} $ & QSO J0831+5245 & 389  &  10908 & 347   &  6213\\
    \hline
    \end{tabular}
    \caption{Average signal and background count rates from the quasars measured during our experiment. All rates are reported in counts per second. While the fluctuations in signal rates varied significantly due to atmospheric seeing, fluctuations in the measured noise rates were consistent with $\sqrt{N}$ error over the course of a 5-minute noise measurement.}
    \label{tab:quasarrates}
\end{table}

\begin{table}[!htbp]
\centering
\begin{tabular}{| c | c | c | c | c |} 
\hline
Pair & $p (a_1)$ & $p (a_2)$ & $p (b_1)$ & $p (b_2)$ \\
\hline
1 & $0.4261$ & $0.5739$ & $0.3505$ & $0.6495$ \\
\hline
2 & $0.5527$ & $0.4473$ & $0.3480$ & $0.6520$ \\
\hline
\end{tabular}
\caption{\small Values for the single-side settings probabilities. }
\label{tab:papb}
\end{table}

\begin{table}[!htbp]
\centering
\begin{tabular}{| c | c | c | c | c | c |}
\hline
Pair & quantity & $11$ & $12$ & $21$ & $22$ \\
\hline
1 & $q_{ij}$ & $0.1486$ & $0.2774$ & $0.2018$ & $0.3721$ \\
  & $p_{ij}$ & $0.1493$ & $0.2767$ & $0.2011$ & $0.3728$ \\
\hline
2 & $q_{ij}$ & $0.1890$ & $0.3637$ & $0.1589$ & $0.2883$ \\
  & $p_{ij}$ & $0.1923$ & $0.3604$ & $0.1556$ & $0.2916$ \\
\hline
\end{tabular}
\caption{\small Values for the joint settings frequencies $q_{ij}$ and inferred joint probabilities $p_{ij}$.
}
\label{tab:qijpij}
\end{table}

We expect that for each dataset, $q_{ij} \simeq p_{ij}$. If we did not find $q_{ij} \simeq p_{ij}$, then (in principle) there could have been some common cause that established correlations among the various setting choices; the choices $ij$ would not be independent. To test for any such violation of independence among the joint settings $ij$, we may conduct a Pearson's $\chi^2$ test by calculating the statistic
%%%%%%%%
\begin{equation}
\chi^2 = N \sum_{ij} \frac{ (q_{ij} - p_{ij} )^2 }{p_{ij} } ,
\label{chidef}
\end{equation}
and then computing the $p$-value corresponding to this value of $\chi^2$ \cite{handsteiner2017a}.
For pair 1, we find $\chi^2 = 0.1504$, which implies that (under the assumption of independent setting choices) there is a purely statistical chance of $p = 0.698$ that the observed frequencies $q_{ij}$ (or data that deviate even more from the inferred $p_{ij}$) would be obtained. For pair 2, we find $\chi^2 = 2.405$, corresponding to a statistical chance $p = 0.121$. Given that each of these $p$ values is larger than typical thresholds (such as $p < 0.05$), we conclude that there is no strong statistical support to refute the hypothesis that the settings $ij$ were independent of each other, for both pairs 1 and 2.

\section{Testing for Violations of No-Signaling}
\label{sec:nosignal}

Another statistical test to consider is whether our data are consistent with the assumption of no-signaling. To be consistent with the principle of no-signaling, the local outcome probabilities must not depend on the setting of the distant detector, which is to say \cite{handsteiner2017a}:
%%%%%%%%%%
\begin{eqnarray}
p (A = + \vert a_i b_j) &=& p (A = + \vert a_i b_{j'} ) , \nonumber \\
p (B = + \vert a_i b_j ) &=& p (B = + \vert a_{i'} b_j ) .
\label{nosignalrequirement}
\end{eqnarray}
The analogous expressions for the `$-$' outcomes follow from $p (A = - \vert a_i b_j) = 1 - p (A = + \vert a_i b_j)$. Point estimates for these conditional probabilities may be estimated from the measured coincidence counts. We define
%%%%%%%%%%%%%%
\begin{eqnarray}
N_{ij}^{A = +} &\equiv& N_{ij}^{++} + N_{ij}^{+ -}  , \nonumber \\
p (A = + \vert a_i b_j ) &\equiv& \frac{ N_{ij}^{A = +} }{ N_{ij}^{A = +} + N_{ij}^{A = - } },
\label{pAijdef}
\end{eqnarray}
with analogous expressions for measurement outcomes $A = -$ and $B = \pm$. Point estimates for the various probabilities are shown in \Cref{tab:nosignalAB}.

\begin{table*}[!htbp]
\centering
\begin{tabular}{ | c | c  c | c  c | c  c | c  c | } 
\hline
Pair  & $p (A=+ \vert 11 )$ & $p (A=+ \vert 12)$ & $p (A=+ \vert 21)$ & $p (A=+ \vert 22)$ & $p (B= + \vert 11)$ & $p (B= + \vert 21)$ & $p (B= +\vert 12)$ & $p (B=+ \vert 22)$ \\
\hline
1 & $0.4201$ & $0.4303$ & $0.5490$ & $0.5559$ & $0.5242$ & $0.5167$ & $0.5936$ & $0.5825$ \\
\hline
2 & $0.4591$ & $0.4534$ & $0.5502$ & $0.5817$ & $0.4996$ & $0.5152$ & $0.5820$ & $0.5976$  \\
\hline
\end{tabular}
\caption{\small Point estimates for the conditional probabilities for measurement outcomes at Alice's and Bob's detectors, with $p (k=+ \vert ij) \equiv p (k = + \vert a_i b_j)$. 
}
\label{tab:nosignalAB}
\end{table*}

We may assess whether these conditional probabilities show any statistical evidence of the violation of no-signaling by performing pooled two-proportion $z$-tests for each relevant pair, such as $p (A = + \vert a_1 b_1)$ and $p (A = + \vert a_1 b_2)$ \cite{handsteiner2017a}, which yields the $p$-values listed in \Cref{tab:nosignalz}. As can be seen in \Cref{tab:nosignalz}, nearly all of the single-sided outcomes from pairs 1 and 2 are consistent with the hypothesis of no-signaling. The only anomalously low $p$-value is for $p (A = + \vert a_2 b_j)$ for pair 2, which yields $p = 0.023 < 0.05$. However, for 8 independent tests, the probability that one shows a $p$-value at least as bad as 0.023 is $1-(1-0.023)^8=0.170$, so the overall suite of tests is consistent with the hypothesis of no-signaling.

\begin{table*}[!htbp]
\centering
\begin{tabular}{| c | c | c | c | c | }
\hline
Pair & $p$ for $p (A = + \vert a_1 b_j)$& $p$ for $p (A = + \vert a_2 b_j)$ & $p$ for $p (B = + \vert a_i b_1)$ & $p$ for $p (B = + \vert a_i b_2)$ \\
\hline
1 & $0.395$ & $0.503$ & $0.562$ & $0.234$ \\
\hline
2 & $0.653$ & $0.023$ & $0.308$ & $0.156$ \\ 
\hline
\end{tabular}
\caption{\small Probabilities $p$ that the observed data (or worse) are obtained under the null hypothesis of no-signaling.}
\label{tab:nosignalz}
\end{table*}

\section{Predictability of Settings}

In this section we consider imperfections in the experiment that can lead to an excess predictability \cite{Kofler16} of the setting choices. Such excess predictability $\epsilon$ quantifies the fraction of runs in which one could predict a specific setting better than would be inferred from the overall bias of the setting choices, given all possible knowledge about the setting-generation process that could be available at the emission event for the entangled particles and thus at the distant measurement events. Loosely speaking, $\epsilon$ quantifies the fraction of runs in which the assumptions of locality and freedom-of-choice fail to hold.

In general, we consider a given trial ``corrupt" if any one of three possibilities occurred at either Alice's or Bob's detector: it (1) involved a noise photon rather than a cosmic photon (where noise photons could arise from either detector dark counts or skyglow), or (2) involved a cosmic photon that was misdirected by a dichroic filter, or (3) involved a cosmic photon that was previewed by the local-realist model for the purpose of considering a dichroic-mirror error, but was passed over because the cosmic photon already had the desired color. 

Given these considerations, we parameterize the excess predictabilities for each detector setting as \cite{handsteiner2017a}
%%%%
\begin{eqnarray}
\epsilon_{a_i} &=& \frac{ 1}{r_{a_i}} \left( n_{a_i} + f_{i' \rightarrow i}^{(A)} s^{(A)} \right) ,\nonumber \\
\epsilon_{b_j} &=& \frac{ 1}{ r_{b_j}} \left( n_{b_j} + f_{j' \rightarrow j}^{(B)} s^{(B)} \right) , 
\label{epsaepsbdef}
\end{eqnarray}
where the rate of detected photons at Alice's detector is given by 
%%%%%%%%%
\begin{equation}
r_{a_i} = (1 - f_{i \rightarrow i'}^{(A)}) s_i^{(A)} + f_{i' \rightarrow i}^{(A)} s_{i'}^{(A)} + n_{a_i} ,
\label{raidef}
\end{equation}
with a comparable expression for Bob's detector, $r_{b_j}$. Here $n_{a_i}$ is the measured rate of noise photons (dark counts and skyglow), which may be quantified by pointing the receiving telescope marginally away from its quasar target; $f_{i \rightarrow i'}^{(A)}$ is the fraction of cosmic photons whose color (if correctly identified) would have led to setting choice $a_i$, but which were misdirected by the dichroic mirror toward the wrong port, yielding setting $a_{i'}$; and $s_i^{(A)}$ is the detected rate of cosmic photons which have a color that (when correctly identified) yield setting choice $a_i$. Because the fractions of misdirected cosmic photons were so small (as noted above), with $f_w < 2 \times 10^{-5}$ for both red-to-blue and blue-to-red at Alice's and Bob's stations, we may neglect the effects from nonzero $f_w$. This simplifies our analysis compared to Ref.~\cite{handsteiner2017a}, in which the various wrong-way fractions were as large as $f_w = {\cal O} (10^{-2})$. For pairs 1 and 2, we find the signal rates and noise rate shown in \Cref{tab:quasarrates}.

As in Refs.~\cite{handsteiner2017a,Kofler16}, we assume that a local-realist model could exploit each ``corrupt" trial so as to produce measurement outcomes that exceed the usual Bell-CHSH inequalities of $C \leq 0$ or $S \leq 2$. In particular, we make the conservative assumption that predictable setting events do not occur simultaneously at both detectors, so that the total fraction of corrupt joint settings is simply the sum of the corrupt settings on each side:
%%%%%%%%
\begin{equation}
\epsilon_{ij} \equiv \epsilon_{a_i} + \epsilon_{b_j} .
\label{epsijdef}
\end{equation}
If any value calculated as in Eq.~(\ref{epsijdef}) exceeds 1, then the corresponding $\epsilon_{ij}$ is set to 1. We further assume that the local-realist model can maximally exploit each corrupted trial, so that the maximum value of $C$ that the local-realist model could attain by exploiting excess predictabilities of detector settings would be \cite{Kofler16}
%%%%%%%%%%%
\begin{equation}
C \leq \epsilon ,
\label{Cepsilon}
\end{equation}
where
%%%%%%%%
\begin{equation}
\epsilon \equiv {\rm max}_{ij} \epsilon_{ij} = {\rm max}_i \epsilon_{a_i} + {\rm max}_j \epsilon_{b_j} .
\label{epsmax}
\end{equation}
Following Refs.~\cite{handsteiner2017a,Kofler16,leung2017}, to ensure
that a sufficient number of genuine quasar photons are detected compared to skyglow, dark current, and misdirected photons, the inequality of Eq.~(\ref{Cepsilon}) places a constraint on the visibility fraction, such that we require 
\begin{eqnarray}
\epsilon & < & \Vexp \sqrt{2} - 1\, 
\label{eq:violation1}
\end{eqnarray}
for all times during the experiment, where the visibility fraction ${\cal V}$ is defined in Eq.~(\ref{Vdef}).
Eq.~(\ref{eq:violation1}), in turn, sets a constraint on the signal-to-noise ratio on each of the four settings ports. Given the values of ${\cal V}$ shown in \Cref{tab:CS}, the constraint of Eq.~(\ref{eq:violation1}) becomes $\epsilon < 0.322$ (pair 1) and $\epsilon < 0.314$ (pair 2). As shown in \Cref{tab:eps2}, all values for the excess predictabilities $\epsilon_{ij}$ easily satisfy this constraint, for both pairs of quasars.

Meanwhile, each of the corrupt fractions $\epsilon_{a_i}$ and $\epsilon_{b_j}$ has some statistical uncertainty, $\sigma_{\epsilon_{a_i}}$ and $\sigma_{\epsilon_{b_j}}$, solely due to fluctuating skynoise during the measurement run. Since the total number of runs is recorded, the only unknown is in the total number of runs that were conducted with noise photons in our CRNGs over our measurement period. This is estimated by measuring the average noise count rates 
before and after the measurement period at each telescope and using the higher of the two count rates to estimate the total number of corrupt runs. In each of our noise measurements we find that the estimated unbiased mean-square error in the count rates is consistent with Poisson noise.

We temporarily drop the labels for Alice and Bob,
and assume that the rates $r_i$ and $n_i$ are independent (which follows from our assumption of fair sampling for all detected photons). Then we find
\begin{equation}
\sigma_{\epsilon_i}^2 = \sigma^2_{n_i} / \bar{r}_i^2 + \mathcal{O}(f_w) .
\label{sigmaepsai}    
\end{equation}
If we further assume that Alice's and Bob's predictabilities are independent, then we find
%%%%%%%%%%
\begin{equation}
\sigma_{\epsilon_{ij}} = \sqrt{ \sigma_{\epsilon_{a_i}}^2 + \sigma_{\epsilon_{b_j}}^2 } ,
\label{sigepsij}
\end{equation}
with an estimated uncertainty on $\epsilon$, as defined in Eq.~(\ref{epsmax}), of
%%%%%%%%%%%
\begin{equation}
\sigma_{\epsilon } = \sqrt{ \sigma_{{{\rm max}_i \epsilon_{a_i}} }^2 + \sigma_{{{\rm max}_j \epsilon_{b_j}} }^2 } .
\label{sigmaepsmax}
\end{equation}
Values of $\epsilon_{a_i} \pm \sigma_{\epsilon_{a_i}}$ and $\epsilon_{b_j} \pm \sigma_{\epsilon_{b_j}}$ for pairs 1 and 2 are shown in Table~\ref{tab:eps1}, and values of the excess predictabilities for joint settings, $\epsilon_{ij} \pm \sigma_{\epsilon_{ij}}$, are shown in Table~\ref{tab:eps2}.

\begin{table}[!htbp]
\centering
\begin{tabular}{| c | c | c | c | c |}
\hline
Pair & \multicolumn{1} {c |}{$\epsilon_{a_1}  \pm \sigma_{\epsilon_{a_1}}$ } & \multicolumn{1}{c |}{$\epsilon_{a_2} \pm \sigma_{\epsilon_{a_2}}$ } & \multicolumn{1}{ c |}{ $\epsilon_{b_1} \pm \sigma_{\epsilon_{b_1}}$} & \multicolumn{1}{c |}{$\epsilon_{b_2} \pm \sigma_{\epsilon_{b_2}}$} \\
\hline
{1} & $0.1441$	&$0.1334$	&$0.0653$		&$0.0342$ \\
	&$ \pm 1.21 \times 10^{-3}$ & $ \pm 0.88 \times 10^{-3}$ & $\pm 0.46 \times 10^{-3}$ & $\pm 0.13 \times 10^{-3}$    \\
\hline
{2} 	& $0.1326$ 				& $0.1679$				& $0.0537$				& $0.0342$ \\
	& $\pm 0.46 \times 10^{-3} $ 	& $\pm 0.54 \times 10^{-3}$ 	& $ \pm 0.93 \times 10^{-3}$ & $ \pm 0.26 \times 10^{-3}$ \\
\hline
\end{tabular}
\caption{\small Values for the fractions of ``corrupt" detector settings at each detector. }
\label{tab:eps1}
\end{table}

\begin{table}[!htbp]
\centering
\begin{tabular}{| c | c | c | c | c |}
\hline
Pair & \multicolumn{1} {c |}{$\epsilon_{11}  \pm \sigma_{\epsilon_{11}}$ } & \multicolumn{1}{c |}{$\epsilon_{12} \pm \sigma_{\epsilon_{12}}$ } & \multicolumn{1}{ c |}{ $\epsilon_{21} \pm \sigma_{\epsilon_{21}}$} & \multicolumn{1}{c |}{$\epsilon_{22} \pm \sigma_{\epsilon_{22}}$} \\
\hline
{1} & $0.2095$	&$0.1783$	&$0.1987$	&$0.1676$ \\
	&$ \pm 1.30 \times 10^{-3}$ & $ \pm 1.22 \times 10^{-3}$ &  $\pm 0.99 \times 10^{-3}$ &  $\pm 0.89 \times 10^{-3}$ \\
\hline
{2} 	& $0.1862$ 	& $0.1667$	& $0.2216$	& $0.2021$ \\
	& $\pm 1.04 \times 10^{-3} $ 	& $\pm 0.53 \times 10^{-3}$ 	& $ \pm 1.07 \times 10^{-3}$ & $ \pm 0.60\times 10^{-3}$ \\
\hline
\end{tabular}
\caption{\small Values for the excess predictabilities for various joint detector settings.
}
\label{tab:eps2}
\end{table}

Since the exact number of runs is known and recorded, we consider the probabilities $\epsilon_{ij}$ to be known (to within some uncertainty $\sigma_{\epsilon_{ij}}$), but the actual number of corrupt trials to be subject to statistical fluctuations. In other words, the occurrence of a corruption in any trial is taken to be an independent random event, which has probability $\epsilon_{ij}$ that depends on the settings pair $ij$. We assume that for ``uncorrupt" trials, the local-realist model has no information about what the settings pair will be beyond the joint settings probabilities $q_{ij}$ \cite{handsteiner2017a,Kofler16}.

In general, the total rates ($r_{a_i}, r_{b_j}$) and noise rates ($n_{a_i}, n_{b_j}$) can vary during the course of an observing period. For example, when observing a given quasar over a substantial period of time, the receiving telescope will collect its light through varying amounts of airmass, as the quasar rises above or moves toward the horizon, thereby affecting the noise rate. In practice, however, our observing periods for both pairs 1 and 2 were brief enough ($\leq 17$ min) that the measured values for $r_{a_i}, r_{b_j}$ and $n_{a_i}, n_{b_j}$ 
did not change substantially; the largest variation among all four detector settings yielded a difference $\Delta \epsilon_{a_1} = 8.6\%$ in the excess predictability. Therefore we adopt the conservative approach of assuming constant values of each $\epsilon_{a_i}$ and $\epsilon_{b_j}$ during a given experimental run, and use the largest values for each detector setting. Such an approach will (modestly) underestimate the statistical significance of our results. 

\section{Statistical Predictability of Random Bits}

Throughout our analysis, we assume that the bits within the sequence gathered from a given quasar are independent of each other. That is, we assume that there is no sequence of bits within the bitstream that contains any information about any future bit. If this independence did not hold, then (in principle) a local-realist mechanism would be able to exploit any excess predictability (beyond the natural bias) to engineer a measured violation of Bell's inequality. Although complete independence among the bits within each bitstream can never be rigorously proven,
we can calculate bounds for the available information.
As we assume fair-sampling for the cosmic photons, we included all detection events, instead of postselecting for those that satisfied the requirements of $\tau_{\rm valid}^k (t)$ to yield a valid setting. For the calculation of the mutual information $\hat{I}(m)$ we adopt the approach developed in Ref.~\cite{leung2017}: We calculate the mutual information between every bit and every sequence of the $m=17$ preceding bits, correcting the biased estimator for the finite size of our bitstream. The value $m$ has been chosen such that it is strictly larger than $\lfloor \log_2(L) - 7\rfloor$ for each measurement file, where $L$ is the total number of detection events~\cite{treves1995upward,rukhin2001statistical}.
The results are presented in \Cref{tab:mutual-info}. The calculated values of $\hat{I} (m)$ are $2-3$ orders of magnitude smaller than the values of the excess predictabilities $\epsilon_{a_i}$ and $\epsilon_{b_j}$ that we already incorporate in our data analysis. These small values of $\hat{I} (m)$ therefore make 
no quantitative impact on our analysis or conclusions.

\begin{table}
\centering
\begin{tabular}{|l|c|l|c|}
\hline
Pair/Side       & Quasar         & $L$ & $\hat{I}(m)$ \\
\hline
1/$\mathcal{A}$ & QSO B0350-073  & 5668580  & $2.0\times 10^{-4}$                          \\
\hline
1/$\mathcal{B}$ & QSO J0831+5245 & 9010082  &   $1.6\times 10^{-4}$                   \\
\hline
2/$\mathcal{A}$ & QSO B0422+004  & 6338028  & $5.1\times 10^{-5}$  \\  
\hline
2/$\mathcal{B}$ & QSO J0831+5245 & 13336320 &   $2.7\times 10^{-4}    $               \\
\hline
\end{tabular}
\caption{We characterize the statistical predictability of our random bitstreams generated by each quasar by computing the mutual information of each bit with the $m=17$ bits preceeding it. This measure corresponds to the probability of guessing the bit correctly given knowledge of the prior 17 bits. The number of lookback bits $m$ for which our estimator is valid is set by $L$, the length of the bitstream.}
\label{tab:mutual-info}
\end{table}

\section{Statistical Significance of Bell Violation}
\label{sec:Bell}

As discussed in Ref.~\cite{handsteiner2017a}, there exist several different approaches to estimating the statistical significance for Bell experiments. The result of any such statistical analysis is a $p$-value, that is, a bound for the probability that the null hypothesis---in our case, local realism with excess predictability of the detector settings $\epsilon_{ij}$, biased detector-setting frequencies $q_{ij}$, fair sampling, fair coincidences, {\it and any other additional assumptions}---could have produced the experimentally observed data by a random, statistical variation.

Until recently, it was typical in the literature to make several simplifying assumptions when calculating the $p$-value for a Bell test, such as that each trial was independent and identically distributed, that the local-realist mechanism could not make any use of ``memory" of the settings and outcomes of previous trials, and that excess predictabilities $\epsilon_{ij}$ could be neglected. Under those assumptions, one typically applied Poisson statistics for single coincidence counts. As emphasized in Ref.~\cite{handsteiner2017a}, however, such an approach assumes that the measured coincidence counts $N_{ij}^{AB}$ are equal to their expected values, only to contradict that assumption by calculating the probability that the $N_{ij}^{AB}$ could have values differing by several standard deviations from their expected values. More recent, sophisticated analyses do not make such assumptions, and represent significant improvements over previous methods \cite{Kofler16,Gill03,Gill14,Bierhorst15,Elkouss16}. However, even these newer approaches are not optimal for our particular experimental arrangement. For example, they are not optimized for experiments with unequal (biased) settings probabilities, and/or they yield non-tight upper bounds for $p$ that can dramatically underestimate the statistical significance.  

For these reasons, in Ref.~\cite{handsteiner2017a} we presented an {\it ab initio} calculation of the $p$-value tailored more specifically to experiments like ours. We follow the same steps here, and refer readers to the more detailed description of the calculation in Ref.~\cite{handsteiner2017a}. First we construct the quantity
%%%%%%
\begin{equation}
W \equiv \sum_{ij} \frac{ N_{ij}^{\rm win} }{q_{ij} (1 - \epsilon_{ij} ) } ,
\label{Wdef}
\end{equation}
with $N_{ij}^{\rm win} \equiv [ N_{11}^{A \neq B} , N_{12}^{A \neq B} , N_{21}^{A \neq B} , N_{22}^{A = B} ]$. If we estimate the conditional probabilities $p (A = B \vert a_i b_j)$ as in Eq.~(\ref{pcorrdef}), then $W = (3 + C) N$, with $C$ defined in Eq.~(\ref{Cdef}). As described in Ref.~\cite{handsteiner2017a}, to avoid the ambiguity that would arise by the need to assume a specific prior probability distribution for the various detector settings, we take the actual number $N_{ij}$ of the occurrences of each settings pair $a_i b_j$ as given. The relevant ensemble with respect to which we calculate probabilities is the ensemble of all possible {\it orders} in which the settings choices could have occurred. The $p$-value we calculate is then the fraction of orderings for which the local-realist mechanism, using its best strategy, could obtain a value of $W$ greater than or equal to the value obtained in the experiment. 

In the absence of noise or errors, the local-realist mechanism could specify which outcomes $(A, B)$ will arise for each of the possible settings $(a_i, b_j)$. The best plans will win for three of the four possible settings pairs, but will lose for one of the possible settings pairs. Therefore a plan may be fully specified by identifying which settings pair will be the loser for a given trial. In the presence of noise and errors, for each time the settings pair is $a_i b_j$, there is a probability $\epsilon_{ij}$ that the trial is corrupt. If the trial is corrupt, it automatically registers as a win. If it is not corrupt, then it has a probability $P_{ij}^{\rm win}$ of registering as a win, where we take $P_{ij}^{\rm win} = p (A = B \vert a_i b_j)$ for $(ij) = (22)$, and $p (A \neq B \vert a_i b_j)$ for the other cases. These considerations motivate the form of $W$ in Eq.~(\ref{Wdef}) \cite{handsteiner2017a}.

The ensemble average of $W$ takes the form \cite{handsteiner2017a}
%%%%%%
\begin{equation}
\langle W \rangle = N (3 + \bar{\epsilon} ) ,
\label{Wavgdef}
\end{equation}
where we have defined
%%%%%
\begin{equation}
\bar{\epsilon} \equiv \sum_{ij} \frac{ \epsilon_{ij} }{1 - \epsilon_{ij} } .
\label{bareps}
\end{equation}
To calculate the statistical significance, we also must calculate $\sigma_W^2 \equiv \langle W^2 \rangle - \langle W \rangle^2$. As in Ref.~\cite{handsteiner2017a}, we calculate $\sigma_W^{\rm opt}$, subject to the condition that the local-realist mechanism may choose the fractions $f_{ij}$ so as to optimize its strategy, where the $f_{ij}$ are defined as the fraction of trials for which the local-realist mechanism chooses settings pair $(ij)$ to be the losing pair. In Ref.~\cite{handsteiner2017a} we found
%%%%%%
\begin{equation}
f_{ij}^{\rm opt} = \frac{1}{2} - q_{ij} + \frac{ N - 1}{2N} \left[ \frac{\epsilon_{ij} } {1 - \epsilon_{ij} } - \bar{\epsilon} \, q_{ij} \right] .
\label{fij}
\end{equation}
Each value of $f_{ij}$ must be non-negative. (If any value is negative, we use a second Lagrange multiplier and re-calculate the $f_{ij}^{\rm opt}$ \cite{handsteiner2017a}.) For pairs 1 and 2, we find all values $f_{ij}^{\rm opt} > 0$ when calculated as in Eq.~(\ref{fij}), so we may use our expression for $\sigma_W^{\rm opt}$ from Ref.~\cite{handsteiner2017a}, namely
%%%%%%%%
\begin{eqnarray}
\left( \sigma_W^{\rm opt} \right)^2 &=& \frac{ N^2}{4 (N - 1)} \left[ \left( \sum_{ij} \frac{1}{q_{ij} } \right) - 4 \right] - N \bar{\epsilon} \nonumber \\
&\quad& \quad + \frac{ N}{4} \sum_{ij} \frac{\epsilon_{ij} }{q_{ij} (1 - \epsilon_{ij} ) }  \\
&\quad& \quad - \frac{1}{4} (N - 1) \bar{\epsilon}^2 + \frac{1}{4} \sum_{ij} \frac{ (N - \epsilon_{ij} ) \epsilon_{ij} }{q_{ij} (1 - \epsilon_{ij} )^2 } .\nonumber 
\label{sigmaWdef}
\end{eqnarray}
We may then calculate the number of standard deviations $\bar{\nu}$ by which the correlations among the measured coincidence counts exceed those that the local-realist mechanism could have engineered by exploiting excess predictabilities:
%%%%
\begin{equation}
\bar{\nu} = \frac{ W - \langle W \rangle }{\sigma_W^{\rm opt} } .
\label{barnu}
\end{equation}

\begin{table*}[ht]
\centering
\begin{tabular}{| c | c | c | c | c | c | c | c | c | c | c | c | c | } 
\hline
Pair & $W$ & $\langle W \rangle$ & $\sigma_W^{\rm opt} $ & $\bar{\nu} $ & $\Delta_\nu$ & $\nu_n$ & $p_{\rm cond}$ & $p_{\textrm{no-m}}$ & $\nu_{\textrm{no-m}}$ & $B$ & $p$ & $\nu$\\
\hline
$1$ & $72\, 224.1$ & $69\, 319.1$ & $290.222$ & $10.01$ & $0.0576$ & $9.46$ & $1.48 \times 10^{-21}$ & $2.96 \times 10^{-21}$ &  $9.39$ & $0.6001$ & $7.41 \times 10^{-21}$ & $9.29$\\
 \hline
 $2$ &  $51\,110.3 $ & $49\, 268.0$ & $242.745$ & $7.59$ & $0.0395$ & $7.30$ & $1.43 \times 10^{-13}$ & $2.86 \times 10^{-13}$ & $7.21$ & $0.5937$ & $7.03 \times 10^{-13}$ & $7.08$ \\
\hline
\end{tabular}\par
\vspace{-0.2cm}
\caption{
\small Quantities relevant to calculating the statistical significance.  }
\label{tab:results}
\end{table*}

The quantities $W$, $\langle W \rangle$, and $\sigma_W^{\rm opt}$ each depend on $\epsilon_{a_i}$ and $\epsilon_{b_j}$ as well as on the coincidence counts $N_{ij}^{AB}$, which we take as given. Whereas Eq.~(\ref{barnu}) takes into account the excess predictabilities, $\epsilon_{ij}$, however, it does not incorporate the uncertainties on the excess predictabilities, $\sigma_{\epsilon_{a_i}}$ and $\sigma_{\epsilon_{b_j}}$. For the next step, we therefore propagate uncertainties on $\epsilon_{a_i}$ and $\epsilon_{b_j}$ to compute the uncertainty on $\bar{\nu}$, which we denote $\Delta_\nu$. The uncertainty $\Delta_\nu$ takes the form \cite{handsteiner2017a}
%%%%%%%%%
\begin{eqnarray}
\Delta_\nu^2 &=& \sum_{a_i} \left( \frac{ \sigma_{\epsilon_{a_i}} }{\sigma_W^{\rm opt}} \right)^2 \left[ \sum_j \frac{ {\cal E}_{ij} }{q_{ij} (1 - \epsilon_{ij} )^2 }\right]^2 \nonumber \\
&\quad& \quad + \sum_{b_j} \left( \frac{ \sigma_{\epsilon_{b_j}} }{\sigma_W^{\rm opt}} \right)^2 \left[ \sum_i \frac{ {\cal E}_{ij} }{q_{ij} (1 - \epsilon_{ij} )^2 } \right]^2 ,
\label{Deltanudef}
\end{eqnarray}
where
%%%%%%
\begin{equation}
{\cal E}_{ij} \equiv N_{ij}^{\rm win} - N \, q_{ij} - \left( \frac{ \bar{\nu} N }{2 \sigma_W^{\rm opt}} \right) f_{ij}^{\rm opt} .
\label{Eij}
\end{equation}
The naive estimate of the number of standard deviations, $\bar{\nu}$ in Eq.~(\ref{barnu}), implicitly assumed $\sigma_{\epsilon_{a_i}}, \sigma_{\epsilon_{b_j}} = 0$, and therefore $\Delta_\nu = 0$. If we allow for an uncertainty in $\nu$ equal to $n$ times the 1-$\sigma$ uncertainty in $\bar{\nu}$, then we should calculate the $p$-value using 
%%%%%%%%
\begin{equation}
\nu_n \equiv \bar{\nu} - n \Delta_\nu .
\label{nun1}
\end{equation}
If we choose $n$ so that $n = \nu_n$, then we find
%%%%%
\begin{equation}
\nu_n = \frac{ \bar{\nu} }{1 + \Delta_\nu} .
\label{nun}
\end{equation}
Assuming a Gaussian distribution for large-sample experiments, we conclude that the conditional probability that the local-realist mechanism could achieve a value of $W$ as large as the observed value $W_{\rm obs}$, assuming that the true value of $\nu \geq \nu_n$, is given by 
%%%%%%
\begin{equation}
p_{\rm cond} = \frac{1}{2} {\rm erfc} (\nu_n / \sqrt{2}) . 
\label{pcond}
\end{equation}
Moreover, if we assume Gaussian statistics for the uncertainty in $\nu$, then there is an equal probability that the true value of $\nu$ is less than $\nu_n$, in which case we must conservatively assume that $W$ might exceed $W_{\rm obs}$. Therefore the $p$-value corresponding to the total probability that $W \geq W_{\rm obs}$ is bounded by 
%%%%%%%%
\begin{equation}
p_{\textrm{no-m}} = 2 p_{\rm cond} .
\label{pnonmem}
\end{equation}
Again assuming Gaussian statistics, we may then calculate
%%%%%%%
\begin{equation}
\nu_{\textrm{no-m}} = \sqrt{2} \, {\rm erfc}^{-1} (2p_{\textrm{no-m}}) .
\label{nudef}
\end{equation}
The subscript ``no-m" stands for ``no-memory," and indicates that these quantities have been calculated without taking into account possible memory effects, which the local-realist mechanism might have been able to exploit. Using the coincidence counts in Eqs.~(\ref{run2coinc}) and (\ref{run1coinc}) and the values of various quantities in Tables~\ref{tab:qijpij}, \ref{tab:eps1}, and \ref{tab:eps2}, we find the numerical values relevant to the calculation of $p_{\textrm{no-m}}$ and $\nu_{\textrm{no-m}}$ as shown in \Cref{tab:results}.

\begin{figure}
\centering
\includegraphics[width=0.49\textwidth]{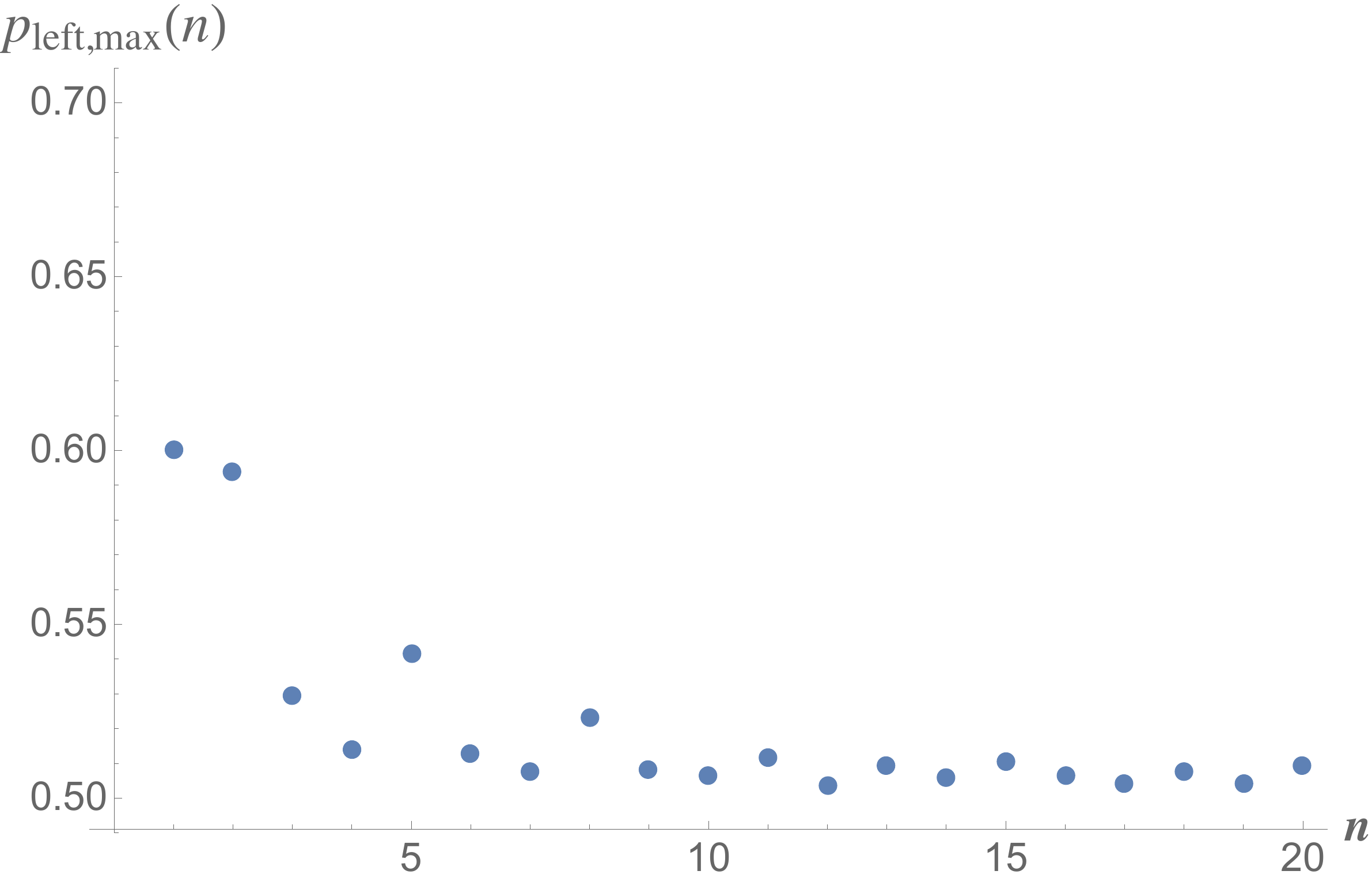} \quad \includegraphics[width=0.49\textwidth]{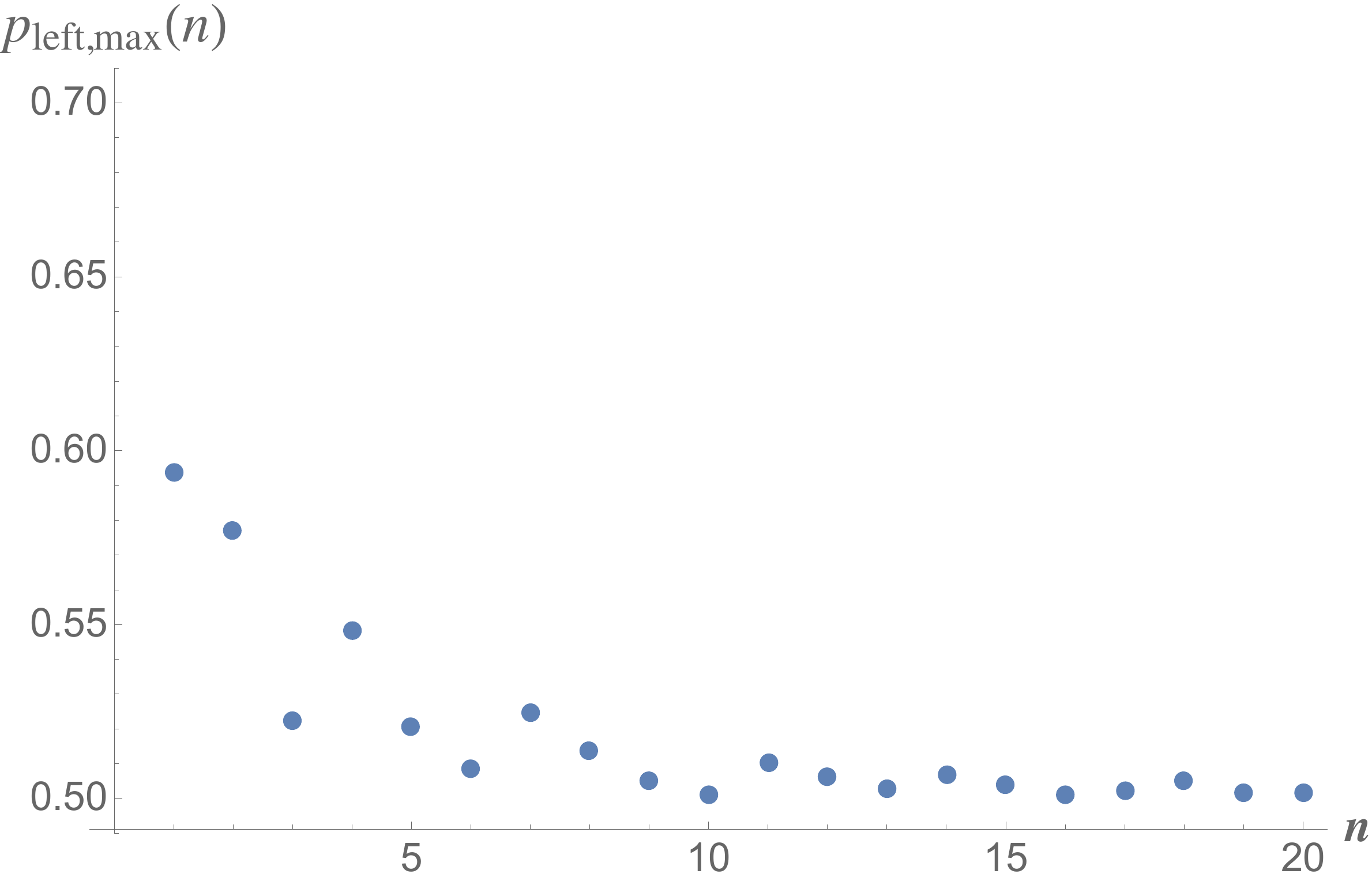}
\caption{The quantity $p_{\rm left, max} (n)$ is the maximum probability that $\tilde{W}$ moves to the left in $n$ trials. Shown here is $p_{\rm left,max} (n)$ for pair 1 (top) and pair 2 (bottom).}
\label{FIG:pleft}
\end{figure}

The expressions for $\langle W \rangle$ and $\sigma_W^{\rm opt}$ in Eqs.~(\ref{Wavgdef}) and (\ref{sigmaWdef}) neglect any advantages that the local-realist model could gain by exploiting memory of previous trials. In the last step of our analysis, we incorporate possible effects from such memory, again following closely the discussion in Ref.~\cite{handsteiner2017a}. We consider the quantity 
%%%%%%
\begin{equation}
\tilde{W} \equiv W - (3 + \bar{\epsilon} ) N ,
\label{tildeWdef}
\end{equation}
with $W$ given in Eq.~(\ref{Wdef}). From Eq.~(\ref{Wavgdef}), we see that $\langle \tilde{W} \rangle = 0$ when averaged over all $N$ trials. Hence the local-realist mechanism cannot change $\langle \tilde{W} \rangle$, but presumably it {\it could} affect the standard deviation of $\tilde{W}$. If we denote by $\tilde{W}_0$ the value of $\tilde{W}$ obtained in the experiment after all $N$ trials, then the $p$-value we seek is the probability that the local-realist mechanism could have achieved $\tilde{W} \geq \tilde{W}_0$ by chance.

We again assume Gaussian statistics for experiments with sufficiently large numbers of trials, $N \gg 1$. Then we expect that as long as $\tilde{W}_n \leq \tilde{W}_0$, the best strategy for the local-realist mechanism is to maximize $\sigma_{\tilde{W}}$, where $\tilde{W}_n$ is the value of $\tilde{W}$ after $n < N$ trials. In this way, the local-realist mechanism would require the smallest number of standard deviations to reach its goal. When and if $\tilde{W}_n$ exceeds $\tilde{W}_0$, on the other hand, then the best strategy for the local-realist mechanism is to minimize $\sigma_{\tilde{W}}$, so as to minimize the probability that $\tilde{W}$ might backslide to $\tilde{W} < \tilde{W}_0$. The quantity we aim to calculate is therefore $p$, which is bounded by
%%%%%%%%%%
\begin{equation}
p \leq \frac{ p_{\textrm{no-m}}}{1 - B} ,
\label{pmem}
\end{equation}
where $p_{\textrm{no-m}}$ is given in Eq.~(\ref{pnonmem}). The quantity $B$ is defined as
%%%%%%%%%%%
\begin{equation}
B \equiv {\rm max}_n \{ p_{\rm left, max} (n) \} ,
\label{Bdef}
\end{equation}
where $p_{\rm left,max} (n)$ is the maximum probability that the quantity $\tilde{W} $ moves to the left after $n < N$ trials. As in Ref.~\cite{handsteiner2017a}, we compute $B$ numerically using the values of $q_{ij}$ and $\epsilon_{ij}$ for pairs 1 and 2. As we had found in Ref.~\cite{handsteiner2017a}, the maximum value of $p_{\rm left,max} (n)$ occurs for $n = 1$. (See Fig.~\ref{FIG:pleft}.) For pair 1, we find $B = 0.6001$, which yields $p \leq 7.41 \times 10^{-21}$, corresponding to $\nu = 9.29$. For pair 2, we find $B = 0.5937$, which yields $p \leq 7.03 \times 10^{-13}$, corresponding to $\nu = 7.08$.


\begin{thebibliography}{86}
\bibitem [{ret()}]{retrocausalnote}%
  \BibitemOpen
  \href@noop {} {}\bibinfo {note} {{One may also consider retrocausal models,
  in which the relevant hidden variables affect the measurement outcomes from
  the future \cite{Beauregard78,argaman10,pricewharton}.}}\BibitemShut {Stop}%
\end{thebibliography}
\end{document}